\documentclass[12pt,onecolumn,showpacs,amssymb,aps,nofootinbib,floatfix]{revtex4-1}
\usepackage{amsmath,amsthm,amssymb,amsfonts,mathtools,centernot}
\usepackage{tensor}
\newcommand{\equref}[1]{Eq.~(\ref{#1})}

\newcommand{\order}[1]{ \mathcal{O} \left( #1 \right) }
\newcommand{\ket}[1]{\left| #1 \right>}
\newcommand{\bfield}{\mathcal{B}}
\newcommand{\efield}{\mathcal{E}}

\newcommand{\bra}[1]{\left< #1 \right|}

\DeclareMathOperator{\Tr}{Tr}
\usepackage{epsfig}
\usepackage{xcolor}
\newcommand{\ave}[1]{\left\langle #1 \right\rangle}

\newcommand{\lnz}{\ln \mathcal{Z}}

\newcommand{\eqcomma}{\phantom{AA},\phantom{AA}}
\newcommand{\braket}[2]{  \left< #1 | #2 \right>  }

\begin{document}
\title{The strong CP problem, general covariance, and horizons}
\author{Giorgio Torrieri, Henrique Truran}
\affiliation{
  Universidade Estadual de Campinas - Instituto de Física ``Gleb Wataghin''\\
  Rua Sérgio Buarque de Holanda, 777\\
  CEP 13083-859 - Campinas SP\\
  torrieri@ifi.unicamp.br\\
}
\begin{abstract}
  We discuss the strong CP problem in the context of quantum field theory in the presence of horizons.
  We argue that general covariance places constraints on the topological structure of the theory.
  In particular, it means that coherence between different topological sectors must have no observable consequence, because the degrees of freedom beyond a causal horizon must be traced over for general covariance to apply.
Since the only way for this to occur in QCD is for $\theta=0$,
  this might lead to a solution of the strong CP problem without extra observable dynamics.
\end{abstract}

\maketitle

\section{Introduction to the strong CP problem}

The strong CP problem \cite{strongcp,qft,shuryak,singular,hooft,witten1979,forkel_instantons,coleman_instantons,adam} is the only fundamental naturalness problem connected to QCD.
It arises generically in Yang-Mills theories as a consequence of the existence of an internal direction in the ``color-symmetry'' gauge group %
$G$, where the vector potential $A_\mu \rightarrow A_{\mu}^{a}$ is allowed to twist in topologically non-trivial ways.
The non-trivial twisting of \( A_{\mu} \) means that finite-action field configurations are separated into topologically separated equivalence classes, labeled by how they twist in gauge space over the asymptotic region \( A_{\mu}(r \rightarrow \infty) \) in Euclidean space \cite{coleman_instantons, forkel_instantons}.
The label for these equivalence classes is an integer \( n[A] \in \mathbb{Z} \), called the winding number of \( A_{\mu} \), given by
\begin{equation}
  \label{winddef}
  n[A] = \frac{1}{32 \pi^2} \int d^4x F_{\mu \nu}^{a} \tilde{F}_{\mu \nu}^{a},
  \quad \tilde{F}_{\mu \nu}^{a} \equiv \frac{1}{2} \epsilon_{\mu \nu \lambda \sigma} F_{\lambda \sigma},
\end{equation}
and all classical solutions of the Euclidean equations of motion of the theory with finite action fall within one of these winding sectors.

Classically, the winding number cannot be changed due to energy conservation.
Quantum mechanically, {however, that can happen due to} the existence of tunneling solutions of the classical Euclidean equations of motion (called instantons) that interpolate between two given equivalence classes \( n \rightarrow n + 1 \), having the form (up to a parameter \( \rho \))
\begin{equation}
  \label{instanton}
A_\mu^a = 2 \frac{x_\nu}{x^2} \left( \frac{\eta^a_{\mu \nu} \rho^2}{x^2+\rho^2} \right),
\end{equation}
where $a$ is the gauge index and $\epsilon_{\mu \nu a}$ mixes internal and spacetime degrees of freedom.

This means the winding number is only conserved perturbatively.
Field strength fluctuations
\begin{equation}
  F_{\mu \nu} = \partial_{\mu} A_{\nu} - \partial_{\nu} A_{\mu}+ [A_{\mu}, A_{\nu}]
\end{equation}
corresponding to field configurations of the type of \equref{instanton} occur locally, suppressed non-perturbatively by the field's action content $\Gamma$
\begin{equation}
  \Gamma \sim \exp \left[ - \frac{1}{4 g^{2}} \int d^{4}x\, F_{\mu \nu}^{a} F_{\mu \nu}^{a} \right],
\end{equation}
and change the field's winding number by one unit.

Note that winding numbers are global properties of given field configurations defined over all space.
Since the topological charge density \( F_{\mu \nu} \tilde{F}_{\mu \nu} \) can be written in terms of a total derivative \( \partial_{\mu} K_{\mu} \) (the so-called Chern-Simons current), they are only ``observable'' (not in practice but with QFT sources, see footnote \ref{footinf}) by probing the celestial sphere at at infinity
\begin{equation}
  \label{windingobs}
  n \sim \int d^4 x F_{\mu \nu} \tilde{F}^{\mu \nu} \sim \lim_{r \rightarrow \infty} \oint d^{3}S \,
  \epsilon_{\mu \nu \rho \sigma}  n_{\mu} \mathrm{Tr} \left(A_{\nu} \partial_{\rho} A_{\sigma} + \frac{2}{3} A_{\nu} A_{\rho} A_{\sigma} \right).
\end{equation}
Instantons, however, are {localized}, and dominate at a scale in momentum space (parametrized by $\rho$ in \equref{instanton}) where the theory becomes non-perturbative.

Instantons break the degeneracy of the classical vacuum solutions with different winding numbers, and their presence means that the Yang-Mills vacuum will be in general a superposition of the different topological sectors weighted by expansion coefficients \( c_{n} \).
  This coefficients are fixed to a phase \( c_{n} = \exp(i n \theta) \) by the homogeneity of Minkowski space \cite{qft}, where the arbitrary constant \( \theta \) parametrizes the coherent superposition of states with topological winding number \( n \) 
\begin{equation}
  \label{qcdvac0}
  Z[\theta] = \sum_{n} e^{i n \theta} Z_{n} \iff \ket{\theta} = \mathcal{N} \sum_n e^{i n \theta} \ket{n}.
\end{equation}
This superposition of winding sectors in the vacuum is equivalent, via imposing the first relation in \equref{windingobs} as a constraint in the partition function, to an additional gauge-invariant \cite{adam} term in the Euclidean effective Lagrangian
%
\begin{equation}
  \label{thetalagr}
  \mathcal{L}_{\text{eff}}
  \equiv
  \ln Z
  =
  \mathcal{L}_{\text{YM}} + \frac{i\theta}{16\pi^{2}} \mathrm{Tr}(F_{\mu \nu} \tilde{F}_{\mu \nu}),
\end{equation}
%
which breaks the CP symmetry of the original Yang-Mills Lagrangian.
In QCD, for example, this CP symmetry breaking gives rise to observable effects, such as $\theta$-dependent electric dipole moments for neutral particles.
Experiments, however, have constrained this CP-violating parameter to the small bound of $\left|\theta \right| \leq 10^{-9}$ (the experimental limit of the neutron's dipole moment is $10^{-18}$ e.m.), beyond naturalness, and the explanation for such a small value is called the strong CP problem.

A variety of solutions \cite{strongcp} were proposed for this problem, most involving beyond the standard model dynamics coupling to $\theta$ (``axions'') or extra symmetries that fix $\theta$ to zero.
Phenomenologically, no signature suggestive of such models has so far been detected.

In this work, we will try to ``go back to the basics'' of quantum field theory and reflect on the nature of the quantum vacuum.
In the following three sections, we will make a series related arguments that link the strong CP problem to the interplay between quantum field theory and general coordinate transformations.
Section \ref{gencluster} argues that a theory with a non-zero $\theta$ will likely lose general covariance of any observables depending on $\theta$.
We shall also show that this constraint on $\theta$ will not affect ``local'' topology fluctuations necessary for 
phenomenology (such as \cite{shuryak,hooft,lattice,instexp,cme}).
Section \ref{caldeira} will show that the ``toy model'' often used to describe $\theta$ vacua together with the tracing over of trans-horizon degrees of freedom leads to $\theta=0$ asymptotic states independently of the ``real'' $\theta$ value.   While our work is new, approaches incorporating some ideas discussed below, such as
\begin{itemize}
\item Constraints from general covariance on local physics \cite{teryaev,sudarsky,matsas,vanzella,gtholo,padma,lovelock,buzzegoli}
\item The coherence of instantons \cite{khleb}
\item Infrared fixed points in the running of $\theta$ \cite{rggroup}
\item The effect of analytical continuation in curved spacetime \cite{etesi,etesi2,blackholes}
\item The stability of periodic states against decoherence \cite{broksym,caldeirabook,caldeira2}
\item The non-invariance of semiclassical states under non-inertial transformations \cite{calixto,mirror}
\end{itemize}
  have been investigated before.   The appendices will also discuss the relationship of our idea to exactly solvable toy models \cite{blomm,blomm2}.
\section{Quantum field theory at different scales: background independence as an infrared symmetry \label{scales}}
At first sight the claim at the end of the last section appears far-fetched.
We are expected to believe that something as ``global'' as the topology of the universe plays an important role in local physics, the effective Lagrangian of QCD\footnote{There is a lot of discussion about weather General relativity is ``Machian'', although Mach's principle played a big role in constructing it.  However the Gibbons-Hawking boundary term looks conceptually like a quantum field version of Mach's principle \cite{gibbons,gibbons2}.  The fact Machian-type reasoning can affect local observables was realized in \cite{teryaev} with regard to the gravitomagnetic moment form factor.}.      
To justify and also sharpen this claim, we need to discuss a little bit more the sensitivity of quantum field theory observables to different scales.    

Analogously to quantum mechanics, in quantum field theory 
all information about the vacuum state is encoded in the generating functional, and perturbations generated by sources can be used to calculate correlations around this state.
In terms of fields $\phi$ and source functions $J(x)$, {the generating functional is given by the expression}
\begin{equation}
  \label{zsourcedef}
  Z = \int [\mathcal{D} \phi] \exp \left[ \int \sqrt{-g} d^4 x \left[ \mathcal{L}(\phi) + J(x)\phi \right] \right].
\end{equation}
For fields of any given spin, both $\phi$ and $J$ acquire the Lorentz and internal symmetry properties necessary for $J(x)\phi$ to be a Lorentz scalar.

Unlike quantum mechanics, quantum field theory \cite{qft} has a continuous infinity of degrees of freedom.
As a result, key theorems behind quantum mechanics, such as the Stone-Von Neumann theorem, stop being valid, {and we are left with the possibility of unitarily inequivalent representations of the canonical commutation relations} (symmetry breaking in field theory is a particularly important example of this).
Mathematically, the construction of observables can be beset by ambiguities, such as divergences.

The main way physicists deal with these issues is in the language of the renormalization group.   Observables in quantum mechanics are not ``states'' but time ordered field correlators of operators $\hat{O}$, $\langle \hat{O}(x_1)\hat{O}(x_2) \rangle$, measured at a scale $Q \sim \left| x_1-x_2\right|^{-1}$.
Thus, unlike in quantum mechanics, even fundamental parameters of the Lagrangian of the theory become ambiguous, related to an observable and sensitive to $Q$.
Provided a stable vacuum is well-defined, 
if one calls the scale of the detector interaction $Q \sim \mu$ and chooses a scale $\Lambda \gg \mu$, one can evolve correlators from $\Lambda$ to $\mu$ and separate the Lagrangian into a finite number of renormalizable terms for which $\Lambda$ can be absorbed into unobservable divergences, and a series in $Q/\Lambda$.  (For correlators, equations of this form are called {Callan-Symanzik} equations \cite{qft}, {while} for Lagrangians {they are called} the Wetterich {equations} \cite{wetterich,polch,schafer}).

This construction has, however, some limitations.  For one, it is only well-defined in Euclidean rather than Minkowski space \cite{turok,nima}, which is obviously an issue when analytical continuation is non-trivial.
As a related point, it has yet not been univocally and generally extended to fully cover {\em infrared physics}, for which $Q$ is much smaller than $\mu$, the inverse of the ``detector size''. Within a locally Minkowski spacetime, this infrared limit could exhibit a non-trivial horizon and causal structure, and evolving from a scale $k \ll \mu$ to a scale $k \sim \mu$ could depend on this. 

For instance, it has long been known that the energy of a vacuum with the boundary is very sensitive to the shape of that boundary, to the effect that it even changes sign \cite{casimir} in ways different from naive dimensional analysis \cite{choi,visser}.  Within the context of the cosmological constant and inflation these topics are subject of active study \cite{choi,desitterinstability,desitterstability}.
While the cosmological constant is a super-renormalizable operator, and dimensionally $\theta$ is superficially a marginal coupling, topological terms are also known to depend on infrared {wavelengths} \cite{renotop}; the relationship with vacuum energy made explicit in color counting \cite{witten1979} and the dependence on the metric topology of \equref{windingobs}.

In {the} case of the $\theta$ vacuum, this mixing between IR and measured {scales} can be seen in the instanton liquid model by calculating the energy expectation value, as was done in \cite{coleman_instantons}
\begin{equation}
  \label{colemaninst}
  <0|H|0>_\theta \sim \int_0 \frac{d\rho}{\rho^5} F \left(\rho \Lambda \right),
  \end{equation}
where $F(...)$ can qualitatively depend on both the measured and IR limit non-trivially.
Physically, this mixing of scales is generated by the fact that these configurations are dominated by high occupation numbers of ground state quanta.  

Since effective field theory techniques rely on a $Q\ll \mu$ expansion, these subtleties are generally incorporated in the $\order{1}$ terms of the effective Lagrangian.   For instance, the calculation of the $\theta$-dependence of the vacuum energy in \cite{ulf} has no trace of the integral in \equref{colemaninst} and appears to only be set by dimensionful parameters of the order of $\Lambda_{QCD}$.   

The arguments above show that, while one must be careful with infrared scales, effective field theories can be a useful guide because they can indicate which terms are compatible with symmetries at the infrared.
For instance, in QED and linearized gravity the fact that emission of quanta of the order $Q \ll \mu$ does not effect observables at $Q \sim \mu$ is related to the symmetries of the theory \cite{weinberg}.  The full consequences of this have been the subject of a lot of theoretical development \cite{nima,weinberg,bms}, and the topology of the space in question (the BMS group in asymptotically flat space) is crucial to this.   

We therefore turn to the other ingredient necessary to construct effective field theories:  Unbroken symmetries in the infrared and in the semiclassical approximation.    The obvious symmetry mentioned previously is the equivalence principle, built out of general covariance of observables.
Extending general covariance to the non-linear gravitational and quantum regime is of course a central, unsolved problem of theoretical physics.
It has long been known that at one loop in the effective theory correlators do not transform covariantly (for a particular example, see the light-bending calculations at \cite{donog}).   It has also been known that quantum vacua with different topologies fall into unitarily inequivalent representations (the ``black hole information paradox'' is a direct consequence of this, as can be seen from \cite{laflamme,gibbons,gibbons2}).

However, from a semiclassical effective theory point of view, the problem does appear more tractable.  It can be argued \cite{gtholo,padma} that if general covariance is to be fundamental one must give up unitarity and treat the degrees of freedom behind the horizon in the language of open quantum systems.   In this case, the vacuum's ``state purity'' is ill-defined, since it is frame {dependent}, but observables could acquire general covariance via the boundary term.   Instead of unitarity, a fluctuation-dissipation theorem would constrain the partition function \cite{zubarev}.
Several lines of evidence point to the fact that such a construction might be achievable.   The soft graviton theorems mentioned earlier \cite{weinberg} relate the independence of infrared physics to Lorentz invariance.
To first order, it has been known for a while that correlators along Rindler paths \cite{laflamme, matsas} transform covariantly once boundary terms are added.
For a correlator $\ave{\psi(x^\mu_1(\tau_1)\psi(x_2^\mu(\tau_2) X}$,where $x_{1,2}^\mu,\tau_{1,2}$ are along a Rindler trajectory and $X$ is either a photon in QED \cite{sudarsky} or a neutrino in Fermi theory \cite{vanzella} one can calculate the Minkowski correlator in terms of interactions with a classical field and the Rindler correlator in terms of interactions with the Unruh bath.
The ``interpretation'' of the calculation will be different but the calculated matrix elements will be the same.

It is therefore worth thinking about the form {that} a ``generally covariant'' interacting quantum field theory would have, and in particular if, in analogy with renormalization requirements, some terms in the Lagrangian would be forbidden by imposing background independence as an infrared symmetry.    In \cite{teryaev}, it was shown, for example, that a non-zero gravitomagnetic moment would violate local covariance of observables under rotations.

In this regard, the $\theta$ term looks suspicious.   The vacuum energy density depends on the boundary and on $\theta$ separately.  The effective $\theta$, $\sim \frac{\delta \lnz}{\delta F\tilde{F}}$ will also depend on the boundary and on the temperature.
Hence, there is no a priori reason for the effective $\theta$ calculated from a Rindler boundary to be the same as the effective $\theta$ at finite temperature, something already clear from group theory arguments \cite{calixto} in a generic scenario with degenerate vacua constructed from conformal zero modes (semiclassical instantons are similar in this respect).
But local Lorentz invariance forces this equivalence \cite{axiom1,axiom2}, leading to non-zero $\theta$ being forbidden.

This suspicious is is complemented, in effective theory language, by the scale separation necessary for both the Unruh effect and instantons to be properly defined. Topological configurations are dominated by high occupation numbers of ground state quanta. In the moving mirror picture mentioned earlier \cite{mirror}, if
one were to construct topologically non-trivial configurations in such a
setup, the relation between Minkowski and mirror boundary conditions
would map ``soft quanta'' carrying topological information into harder
ones which, because of asymptotic freedom, are insensitive to the
presence of instantons in the vacuum.

Acceleration and force are by their nature semiclassical concepts, since they consider momentum a differentiable number rather than an operator.   This means that acceleration's lifetime $T$ needs to be long-lived compared to its scale $a$, and also smaller than the fundamental scale of the detector, for example {its} mass $M$
\begin{equation}
\label{hyerarchy}
 1/T \ll a \ll M  \underbrace{\Rightarrow}_{instanton} \mu \ll a \ll \rho
\end{equation}
in this context, the instanton's bulk of the action is in the peak, of size $\sim \rho$ (\equref{instanton}), yet the topology information will be in the tail, dominated by a diverging occupation number of quanta of characteristic frequency $\ll 1/\rho$.   One can therefore have an acceleration small enough w.r.t. the instanton size, long enough to maintain the semiclassical approximation, where the semiclassical expansion will break down at the tail.   The first hierarchy in \equref{hyerarchy} is therefore far more dubious in its applicability than the second\footnote{One can choose a gauge, such as the ``singular gauge'' \cite{singular}, where the Winding number is not asymptotic, at the price of having a singularity in the center, $A_\mu \sim \partial_\mu \ln \left(1+\rho^2(x-x_0)^{-2} \right)$.  However, the continuation of this Gauge in Minkowski and Rindler space, discussed in the next section as well as \cite{etesi}, is probably impossible for obvious reasons}.

This insight can be sharpened in lower-dimensional theories, where the non-trivial interplay of asymptotic global symmetry properties of the theory and topological terms is well-known \cite{dunne}.   
Appendix \ref{ring} describes in detail a quantum particle on a ring with a magnetic field in thermal equilibrium.  The different geometric and thermodynamic temperature in this system can be mocked up by a device that changes the magnetic field based on the system's heat capacity.    While such a setup can be engineered to add no entropy (no microstates are measured, and the adiabatic limit is maintained) in this case, the thermodynamic temperature is different from the geometric temperature (given by the time periodicity) by a generally non-perturbative factor, \equref{gens}.
  Note that the difference is not necessarily connected to a scale separation, since the two dimensionful scales combine non-perturbatively.

Section \ref{maxwell} summarizes an equivalent 1+1 dimensional topological theory.   As shown there, while in such a theory the $\theta$ term is set as a boundary condition, a source ``communicating'' with a boundary could modify the temperature from the Unruh value by arbitrary amounts.    Dynamics spontaneously creating such topological terms in 3+1 theory, therefore, could potentially mean that the geometric temperature of an accelerating observer would be different from the temperature the observer would measure from the background.
A dynamical symmetry breaking would be equivalent to assuming an effective $\ave{J_L}$  in \equref{blomsource}, coinciding with that of \equref{windingobs}.
As the next section \ref{gencluster} will argue, unless $\theta=0$ it is impossible to make such a term truly background independent.  Section \ref{caldeira} will further show, using the often-used periodic potential analogy, how the IR tracing out could wash out an arbitrary $\theta$ to an effective $\theta=0$.

Summarizing, a ``background-independent quantum field theory'''s requirement would be the separation of local physics (at scale $\mu$) from scales sensitive to the topology of space.     This is very different from saying the latter are relevant to observable physics (indeed, as \cite{ulf} shows explicitly, they are not).
Analogously to counter-terms in Wilsonian renormalization enforcing independence of detectable physics from its UV completion, infrared constraints enforce independence of local physics from the topological features of the chosen coordinate system.
{One such constraint would be \( \theta = 0 \).}

\section{The topological term from the partition function \label{gencluster}}
As can be seen in the previous section, 
the question of the general covariance of quantum field theories is a subtle one.
It is {straightforward} to write an action invariant under general coordinate transformations, since $\sqrt{-g} d^4 x$ is generally covariant and $\mathcal{L}$ can be made so by the use of covariant derivatives.  Indeed, a $\theta$ term can also arise within an effective theory extension of general relativity \cite{deser}.

The issue is the role of the integration measure $ \mathcal{D} \phi$.
What configurations are counted if the coordinate system contains singularities, or the spacetime is divided into {causally} disconnected regions?
We note that, as shown in \cite{hawkingano}, Hawking radiation can be thought of as an anomaly, i.e.\ a tension between the fundamental symmetry of the theory and the integration measure -- in this case provoked by the boundary structure of the Schwarzschild spacetime at the horizon.

At the partition function level, it has long been known \cite{gibbons,gibbons2} that the partition function for general coordinate systems with causal horizons will necessitate of a surface term, which for a {timelike} Killing horizon can be modeled by a thermal bath.
This is the essence of the Unruh effect, and, in 1+1D, can be rigorously connected to topological terms (see appendix and \cite{blomm}).

In \cite{gtholo}, it was proposed that perhaps promoting the partition function with a source to a dynamical object would generate a generally covariant quantum theory.
Since bulk general relativity is {\em always} holographic \cite{padma} (something that can be seen as a consequence of Lovelock's theorem \cite{lovelock}),
a general non-inertial transformation will alter both bulk and boundary, but there is a possibility that, with Lagrangians describing both, the total partition function will be invariant under general coordinate transformations.
In \cite{gtholo} we argued that imposing this must lead to treating all quantum states as open, since such general coordinate transformations necessarily break unitarity.

This extension, however, is unobservable in experiments done so far as it concerns situations with strong classical accelerations.
Such an approach, as we argued, could be used to write down an effective quantum field theory covariant under general coordinate transformations.
Not quantum gravity, of course, but a field theory respecting the symmetries of gravity at quantum level.

In this spirit, let us try to define $\theta$ as a running parameter.    We immediately see that \equref{thetalagr} and \equref{windingobs} are only valid in flat Euclidean space, but any infrared subtlety, like the presence of a horizon, would change the asymptotic shape of the instanton and hence the $k \ll \mu$ dynamics.    Put it differently, while of course $F_{\mu \nu} \tilde{F}^{\mu \nu}$ is a generally covariant scalar the equation \ref{windingobs} does not transform covariantly and hence the constraint leading to \equref{thetalagr} is not generally covariant.

We can however circumvent this problem remembering the effective action can be defined also in terms of a source in \equref{zsourcedef}.
The winding number will be related to 
the infrared limit of a probe such as a ``loop'' of ``color sources'' $J_a^\mu$ placed at infinity\footnote{\label{footinf}A technical note: of course it is impossible even in principle to put a probe at infinity.   However, one can define such a measurement in terms of Bayesian limits.   Here, by ``a probe at infinity'' we mean a probe placed at a sequence of larger and larger distances.  The chance of the winding number not being inferred correctly by the measurement goes down with distance in a calculable way.  Such a sequence can be used to define something like \equref{windingobs}}, while $\theta$ will be related to the entanglement of such winding numbers
\begin{equation}
  \label{windingobs2}
  n \propto \lim_{r_0 \rightarrow \infty} \mathrm{Tr} \oint dS_{d-1}.J^a \delta(r-r_0) A_a  \eqcomma \theta \propto -i\ave{n}
\end{equation}
however, in Minkowski space $S_{d-1}$ can be oriented in either a time-like or a space-like direction.   In the first case, we will be projecting on $\theta$, as it commutes with the Hamiltonian and relation \equref{windingobs} holds.
As a particular case,  if instead of an isotropic $dS_{d-1}$ we consider a ``long thin sausage'', our observable coincides with the dipole moment of a neutral particle antiparticle combination.  This measurement is in fact ``asymptotic'', since observing the neutron for any amount of finite time there is a finite probability that topological fluctuations will give us an effective non-zero dipole moment (of course this is irrelevant for the macroscopic scales where the neutron dipole moment is measured).   

 In the second, spacelike loop, we will be sensitive to the topological number $|n>$ of our system, rather than $\theta$.     This ambiguity is unique for ``topological terms'', and suggests that general topological terms are {\em not} generally covariant.     Indeed,
 it is obvious that the boundary integral has the same structure of equation (\ref{windingobs}); A function of $A_{\mu}^a$ integrated over the surface $dS_{d-1}$ which in turn is the edge of a particular geometry.
A non-zero $\theta$ term means that different $n$'s are coherently entangled, and the degree of this coherence would be modified by the boundary term.
Hence, the entanglement entropy, {\em and all of its derivatives} ($\theta$ is proportional to the first derivative, the topological susceptibility to the second), would be modified by the boundary term in a way that is very sensitive to the geometry in question.
In fact, calculating \equref{windingobs} using the WKB approximation \cite{parikh} and a path crossing the horizon can easily be seen to be boundary-dependent.

This suggestion has been put on a firmer footing within \cite{etesi}, following an unsuccessful attempt by the same author \cite{etesi2} to argue, in a manner similar to \cite{blackholes}, that Yang-Mills instantons are incompatible with a Schwarzschild geometry.
As it turns out \cite{etesi} this is not quite correct.
However, unlike in Minkowski the instanton in Euclidean Schwarzschild space cannot be gauge transformed into a smooth temporal gauge; By looking at a Rindler patch of this space, it is clear that a Minkowski and a Rindler observer, or a Schwarzschild vs a Lemaitre observer, will see a different instanton content and a different $n$.
As $n$ is a scalar, the only way to preserve an undetermined $n$ (required by local quantum mechanics) that transforms covariantly is to make sure the summation over $\left|n\right>$ was incoherent even in the freely falling frame, where geometry is closest to Minkowski.
This corresponds to the case of $\theta=0$.
Note that this argument only applies to the {\em average} topological value $\ave{F \tilde F}$.
Phenomenologically useful fluctuations $\sim \ave{F \tilde{F} F \tilde{F}}$ \cite{lattice} are not affected because of the time-ordered nature of the product in the fluctuation \cite{witten1979}, which isolates the {\em local} (instanton peak, always in causal contact with the observer) over the global asymptotic state (possibly affected by the horizon).  In a Euclidean spacetime, without horizons, the infrared limit of  $\ave{F \tilde{F} F \tilde{F}}$ is related to $\ave{F \tilde F}$
\begin{equation}
  \label{euclmink}
  \lim_{k\rightarrow 0} \int d^4 x e^{ik(x-x')} \ave{F \tilde{F}(x') F \tilde{F}(x)} \sim\ave{F \tilde F}  
  \end{equation}
but in locally Minkowski spacetime, whose global causality structure (set of points where $k_\mu (x-x')^\mu=0$ of \equref{euclmink}) is non-trivial, the two could be different (See the discussion about time orderings in \cite{witten1979}).
The analogy here is the cosmological constant (of which $F \tilde{F}$ is the ``topological'' QCD part, \cite{witten1979}) and local gravitational physics.  Background independence requires that infrared physics could get corrections from quantum fluctuations and decoherence \cite{choi}, but the same symmetry requires that locally physics is unaffected.  This can happen if all local physics is sensitive to the second derivative and the first derivative is zero, as indeed seems to be the case.

Let us explore this argument in more detail, but concentrating on Rindler patches. 
For static and quasi-static spacetimes this is equivalent to filling the manifold with a bath of quanta with the temperature $T_h \sim 1/R_h$, where $R_h$ is the horizon scale (the Schwarzschild radius for black holes, the Hubble radius for dS space, the acceleration for Rindler space and so on).
Thermality appears as a consequence of the symmetries of the quasi-static accelerated spacetime \cite{witten,calixto}, and hence can be thought of as as embedded in the {\em effective action}
\begin{equation}
  \label{seff}
  S_{eff} \sim \ln\left[ \int \mathcal{D} \phi \exp \underbrace{\int \sqrt{-g} d^4 x}_{\rightarrow \oint dt \int d^3 x \sim T \sum_n \delta\left( t- \frac{n}{T} \right)} \left[ \mathcal{L}(\phi) + J(x)\phi \right] \right].
\end{equation}
As shown in \cite{axiom1,axiom2}, for axiomatic field theory (defined in terms of correlators), Lorentz invariance implies this relation is exact.

As further shown in \cite{matsas,laflamme} and references therein, one can derive the Unruh effect by tracing over degrees of freedom beyond the horizon.
\begin{figure*}[h]
  \epsfig{width=14cm,figure=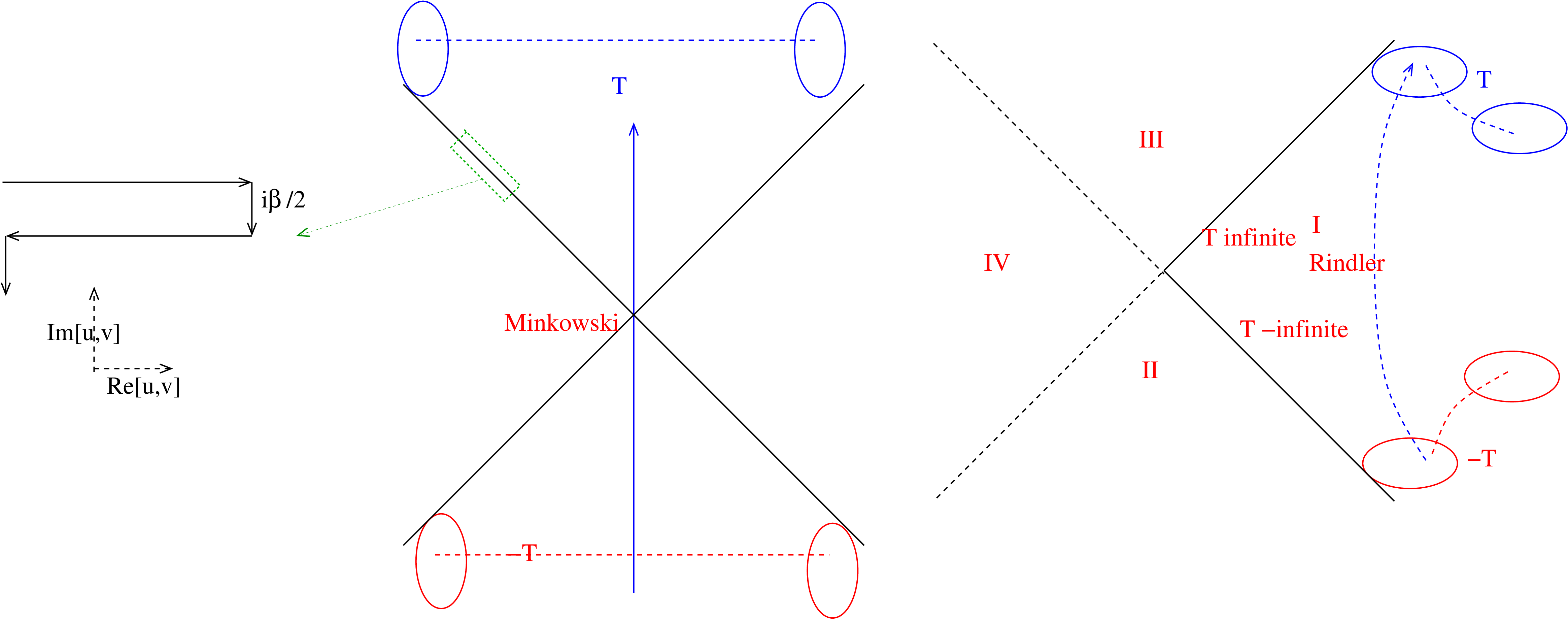}
  \caption{\label{rindler}
    The asymptotic horizons of Minkowski and Rindler observers, on which winding numbers are measured (middle and right panels).
    As shown in \cite{laflamme}, a rigorous dictionary between the two partition functions can be established by complexifying coordinates and a choice of contours, shown on the left
  }
\end{figure*}
The two pictures are complementary since, at least for Rindler patches \cite{horibe,truran}, tracing over in Minkowski space can be achieved by complexifying the action and appropriately choosing contours constructed to respect the periodicity of the time coordinate.
This way we can write, up to a normalization factor, the generating functional for a field as
\begin{equation}
  \label{eqn:generating-functional}
  Z[J]
  \sim
  \int \mathcal{D}_{\mathcal{C}} \phi \exp
  \left\{
    i \int_{\mathcal{C}}
    \mathop{\sqrt{-g} d^{4}x}
    \left(
      \mathcal{L}[\phi, \partial \phi]
      + J \phi
    \right)
  \right\},
\end{equation}
where \( \mathcal{C} \) corresponds to a contour choice in complex coordinate space -- analogous to the Schwinger-Keldysh and thermofield-dynamical formalisms of usual finite-temperature quantum field theory.
An extension of Rindler to Minkowski spacetime can be realized, in Rindler null-coordinates \( l_{\pm} = \tau \pm \xi \), by the horizontal patches \( \mathcal{C}_{1}, \mathcal{C}_{2}\) of the contour shown in figure (\ref{rindler}), defined by
\begin{equation}
  \mathcal{C}_{1} \equiv l_{\pm},
  \quad
  \mathcal{C}_{2} \equiv l_{\pm} - \frac{i \beta}{2},
  \quad
  l_{\pm} \in \mathbb{R}.
\end{equation}
In this way, all four  wedges \( W_{\pm,F,P} \) of Minkowski space correspond to different combinations of horizontal sections of the Rindler null-coordinate contours.
The fields associated with the horizontal sections \( l_{\pm} \in \mathcal{C}_{2} \) (with non-zero imaginary part \( \Im \mathcal{C} \neq 0 \)) correspond to the causally inaccessible regions of spacetime \( W_{-,F,P} \), and act as invisible fields from the point of view of an accelerated observer.
In other words: The extension of the fields from Rindler to Minkowski spacetimes acts as a purification of the Hawking-Unruh thermal state\footnote{
{Contour choice has also been contentious in the question of whether de-Sitter space evaporates or not. The instability of de-Sitter space was argued for in \cite{desitterinstability}, while \cite{desitterstability} argues de Sitter space is stable. At the heart of the disagreement is the contour definition, with \cite{choi, desitterinstability} relying on a Feynman type locally causal quantization, and the definition of the ``stable'' Bunch-Davies vacuum \cite{desitterstability} relying on a time-symmetric contour constructed not to evaporate. This work takes the first of these two approaches.}}.

Extending such a calculation to a topologically non-trivial Yang-Mills theory is a formidable project. 
However, the arguments in \cite{etesi} make it clear that the fact that winding numbers can be both in the observed and the hidden patches will generally change the topological structure of the resulting partition function.
A physical reason is that for real positive quark masses, the partition function in equation (\ref{qcdvac0}) admits a quasi-probabilistic interpretation
\begin{equation}
  \label{sumnz}
  Z = \sum_n Z_n e^{i n \theta}, \quad P_n \sim \frac{Z_n}{Z(\theta=0)},
\end{equation}
with $P_n$ is the ``probability'' to ``measure'' a winding number $n$ \cite{ulf}.
Note that \cite{ulf} this is a Wigner Quasi-probability \cite{wigner} rather than a probability, since for generic $\theta$ it might not be real-valued.
However, it expresses the quantum uncertainty of winding numbers when $\theta$ is fixed, and it obeys Wigner's quasi-probability axioms.
The winding number, in this picture, is measurable via the probe in equation (\ref{windingobs}).

In the quasi-probabilistic interpretation motivated in equation (\ref{qcdvac0}), $Z_n$ and the topological susceptibility $\partial^2 \ln Z/\partial^2 \theta$
will transform between Minkowski and Rindler space with a factor representing the ratio of visible to invisible fields.
By Bayes's theorem
\begin{equation}
  \label{pnh}
  Z_n =
  {Z(\theta=0)} \sum_{\mu} \frac{Z_{\mu}}{Z(\theta=0)} \times P\left( m - n \in \mathcal{C}_2 \right),
\end{equation}
and the $P(...)$ term will be directly proportional to the proportion of the given Rindler time-slice covering each section of Rindler space.
This is not unity, and will depend on the proper time of the Rindler observer.
It will also be an observable, measurable in a Gedankenexperiment by repeated applications of the operator defined in equation (\ref{windingobs}).

This lack of covariance has a root in two issues: the winding number is not a conserved quantum number, and hence is expected to change with the Hamiltonian.
However, ``time'' is not just a coordinate but also defines the order in which the co-moving observer makes observations (``collapses the wavefunction'', or, rather, samples correlators).
If one sequentially ``observes it'', over time intervals of the order of an instanton size, one expects it to change by one unit.
However, this change in Hamiltonians will also affect the partition function according to
\begin{equation}
  Z
  =
  \bra{0} \exp \left[i \int dt \hat{H}\right] \ket{0},
\end{equation}
and so the procedure \[\ \int dt \rightarrow \frac{1}{N} \sum_n \int_{nT/N}^{(n+1)T/N}dt \] would break down.
If the winding number is observed along a Minkowski vs.\ a Rindler trajectory, the degree of quantum coherence, and $P(...)$ will vary.
The only way to make equation (\ref{pnh}) generally covariant appears to be for $P\left( m - n \in \mathcal{C}_2 \right)$ to be a unity operator in the winding numbers basis, which is equivalent to assuming $\theta=0$.

Note that, since we are describing the vacuum rather than correlators, axiomatic field theory results regarding the equivalence of thermal and accelerated dynamics \cite{axiom1,axiom2} need not apply.  In fact, symmetry arguments can be used to understand how  non-perturbative vacuum degeneracies change the Unruh state w.r.t.\ a thermal state \cite{calixto} (note that the $\theta$ state there is not a QCD one but a generic zero mode condensate).  The result of \cite{calixto} and the qualitative discussion here would imply the accelerated effective $\theta$ is not the same of the effective $\theta$ at finite temperature.

To get a physical feeling of what is going on here, consider the case of accelerated photon ``Bremsstrahlung'' emission, or the famous $p\rightarrow n e^{+} \nu$ decay examined in \cite{matsas}.
As \cite{matsas} makes the case, the ``interpretation'' in inertial and comoving frames is different (in one case the decay is a quantum reaction to a semiclassical field, in the other it is interaction with the Unruh bath) the decay matrix elements calculated in both cases will be the same.  Neutrino oscillations somewhat complicate this last point, and there is no consensus to resolve this\footnote{In particular \cite{neutmatsas} argues that general covariance implies fundamental states mass be the usual mass-shell irreducible representations of the Lorentz group, while \cite{neutblasone,neutcp} argues that on the contrary flavor states are fundamental, with the latter paper showing CP violation would generate an extra violation of general covariance in the mass basis. Finally \cite{neutus} argues that a generally covariant effective action, taking condensates into account, is necessary to resolve the issue.}.

The equivalent Gedankenexperiment would have us compare the EDM measured on an {accelerated} neutron, interpreted by a Minkowski observer (who sees the effect of {acceleration}) and the comoving observer (who sees a finite temperature $\theta$).  Because the causal structure of the two observers are different, so the EDM operator sampled by \equref{windingobs2} will contain a different combination of winding numbers and $\theta$.
{In fact, the violation of the equivalence principle at finite temperature, long known and recently calculated in \cite{buzzegoli}, makes it quite likely from thermal considerations alone.}
Hence, a Minkowski observer and a comoving observer will see different $Z_n$, and hence different effective $\theta$'s and topological susceptibilities.
{\em The only way general covariance is to be a fundamental principle}, $\theta$ must be equal to zero, so the sum of different topological sectors appears incoherent in all frames.

This looks quite an abstract argument, but it has been known for some time in the context of lower dimensional Chern-Simons theories, known as the ``finite temperature puzzle'' (see section 5.4 of \cite{dunne} and references therein), {where finite temperature breaking of Lorentz invariance introduces violations of large Gauge invariance order by order}.  In \cite{dunne}, the breaking of Lorentz invariance is physical, because the system is prepared at finite temperature.   However, such a ``finite temperature'' state could never describe the ground state from the vantage point of a reference frame.

The above discussion can also be incorporated into the effective theory \cite{consti} because in such local models ``local'' and ``large'' Gauge transformations are separated, and the former handled perturbatively by adding a Wilson line $U_\infty$ at infinity to each color charge.
This way, in the perturbative limit, topological transformations completely decouple from local transformations and only the latter are relevant for Feynman diagram expansion.
The problem is that once a horizon exists, $U_\infty$ will span both ``visible'' and ``invisible'' fields.
Hence, no effective theory, perturbative or otherwise, can be made from visible fields alone and topological and local transformations are inseparable.
Naively we can speculate that since (as argued in \cite{consti}) local color-charged objects are forbidden by confinement, this means the $\theta$ angle should also go to zero. Note that the association between confinement and $\theta$ can be made using renormalization group arguments \cite{rggroup}.  Of course, as argued in section 4.1 of \cite{polch,schafer}, there is a case for relating these arguments, with renormalization scheme independence taking the role of general covariance.   If one wants, the symmetry of general covariance and the arguments in section \ref{scales} provide a theoretical justification for the infrared fixed point to be of the form of \cite{rggroup}.

To summarize this section, we made a heuristic and speculative argument that the only way to restore general covariance is for $\theta$ to be equal to zero.
{The crux is that only for $\theta=0$ the coefficient $c_m=e^{im\theta}$ does not change between a coherent and an incoherent sum.
}In the next section we shall show, using a toy model from condensed matter physics solved in the 80's \cite{caldeirabook,caldeira2}, that explicitly tracing over degrees of freedom beyond the horizon will generally confirm this conclusion.

\section{The Periodic potential wells picture of $\theta$ \label{caldeira}}


Since it is fashionable to illustrate the $\theta$ problem using the periodic potential well \cite{shuryak,forkel_instantons,coleman_instantons}, let us try to get some additional understanding using such a toy model.
The models are physically completely different, in the sense that the periodic potential well is ``engineered'' to have a $\theta$-like parameter characterizing the eigenstates of the Hamiltonian and topological terms have no kinetic modes.
Nevertheless, we think that a dictionary between these problems is useful enough to extend it to an open quantum system.
\begin{figure}[t]
  \epsfig{width=18cm,figure=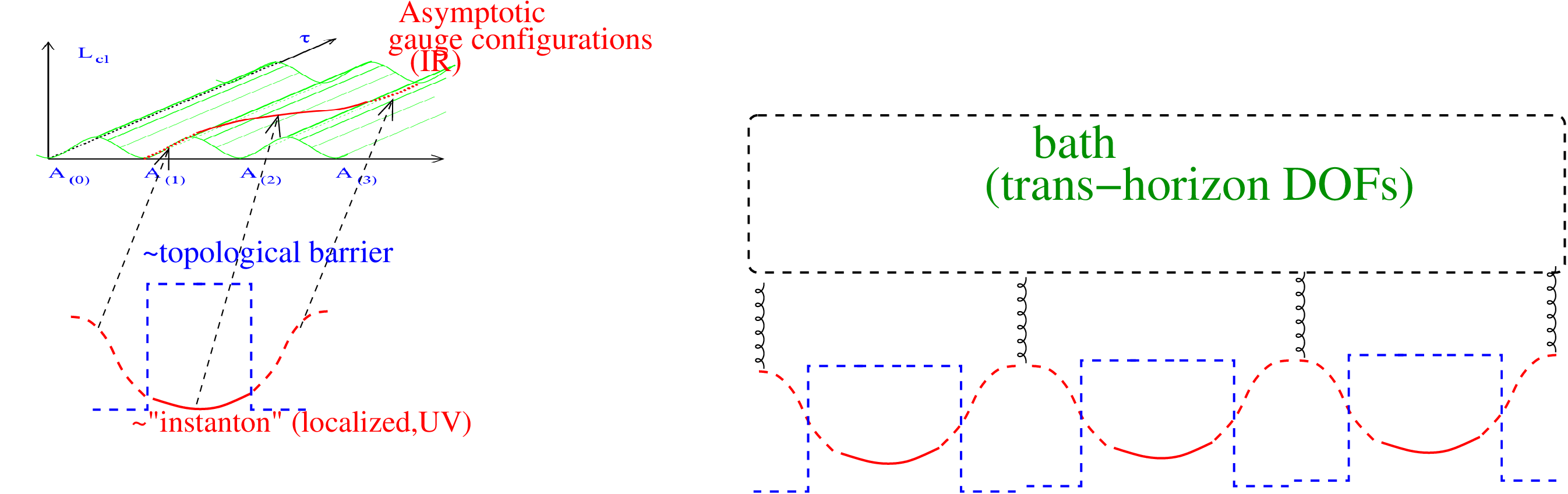}
  \caption{\label{wells}
    A schematic illustration of the periodic well analogy of instantons, and the role that horizon radiation plays in them.
    The QCD solution is represented by a periodic potential, whose wavefunction is represented by a tunneling process (corresponding to a localized field configuration, the instanton) and a dominant ``peak'' (corresponding to the asymptotic field configuration, delocalized).
    Horizon radiation decoheres the IR peak without altering the localized tunneling.
    Top image from \cite{forkel_instantons}
  }
\end{figure}
Let us consider a system consisting of a quantum mechanical particle moving through potential wells satisfying the periodicity condition \( V(x \pm a) = V(x) \), shown by the black lines in Figure (\ref{wells}).
The wells, in this example, represent topological configurations (winding numbers) of a non-abelian gauge theory at the horizon.
The Hamiltonian for this system is given by the kinetic and potential terms
\begin{equation}
  \label{hdef0}
  \hat{H}_{S}
  =
  \frac{\hat{p}^{2}}{2M} + V(\hat{x}).
\end{equation}
The solution for these kinds of systems is well known \cite{sakurai}, and given in terms of the energy eigenstates
%
\begin{equation}
  \label{psicoh}
  \ket{\theta}
  =
  \sum_{n \in \mathbb{Z}} \exp(i n \theta) \ket{n},
\end{equation}
where \( \ket{n} \) is state of the particle localized at the \( n \)the site (or winding number, in the Yang-Mills correspondence), and \( \theta \) is a parameter that labels the simultaneous eigenstates of the Hamiltonian and the \( a \)-translation operator \( T^{\dagger}(a) V(x) T(a) = V(x + a) \).
The associated  density matrix for such $\theta$ state is given by
\begin{equation}
  \hat{\rho}
  =
  \sum_{m,n \in \mathbb{Z}} \exp(i (m - n) \theta) \ket{m} \bra{n}.
\end{equation}
If the potential barriers are tall with respect to the energy of the system, we can use the tight-binding approximation and obtain an approximation to the \( \theta \)-state energy%
%
\begin{equation}
  E(\theta) = E_{0} - 2 \Delta \cos \theta
\end{equation}
from the energy of the localized states \( E_{0} = \bra{n} \hat{H} \ket{n} \) and the splitting energy between adjacent sites \( \Delta = |\bra{n \pm 1} \hat{H} \ket{n}| \).
The ground-state wavefunction \( \psi_{\theta}(x) = \braket{x}{\theta} \) takes the Bloch form
\begin{equation}
  \psi_{\theta}(x)
  =
  e^{i k x} u_{k}(x), \quad k = \theta / a,
\end{equation}
where \( u_{k} \) a periodic function with period \( a \) and \( k = \theta / a \) is the ``Bloch momentum'' associated with the periodic potential.

We now need to add to the toy model of the $\theta$ vacuum a toy model of the thermal bath due to the horizon, and the interactions of the system with the bath.
Following \cite{caldeirabook,caldeira2}, let us add a ``bath'' consisting of an infinite number harmonic oscillators interacting with the particle in the periodic potential as
%
\begin{equation}
  \label{hdef}
  \hat{H} = \hat{H}_{S} + \hat{H}_{B}
\end{equation}
%
where \( \hat{H}_{B} \) is the Hamiltonian for the simple harmonic oscillators of the thermal bath together with a linear interaction term between the bath and the particle, given by
\begin{equation}
  \hat{H}_{B}
  =
  \sum_{n \in \mathbb{Z}}
  \left( \frac{1}{2 m_{n}}\hat{p}_{n}^{2} + \frac{1}{2} m_{n} \omega_{n}^{2} \left( \hat{q}_{n} + \frac{C_{n} \hat{x}}{m_{n} \omega_{n}^{2}} \right)^{2} \right).
\end{equation}
For convenience, we define the pure interacting Hamiltonian by \( \hat{H}_{I} = \sum_{n \in \mathbb{Z}} C_{n} \hat{q}_{n} \hat{x} \).

In the interaction picture, the equation of motion for the reduced density matrix of the particle is given by
%
\begin{equation}
  \frac{d \hat{\rho}_{S}}{dt}
  =
  - i \Tr_{B} \left[\hat{H}_{I}(t), \hat{\rho}_{S + B}(t)\right].
\end{equation}
When the couplings \( C_{n} \) are equal to zero one recovers equation (\ref{psicoh}) for the ground state of the particle, since \( \hat{\rho}_{S} \) and \( \hat{\rho}_{B} \) decouple and evolve, respectively, under \( \hat{H}_{S} \) and \( \hat{H}_{B} \) separately.
If \( C_{n} \) are not equal to zero, the toy model can only be solved in very particular cases and under certain approximations.
Analytical solutions can be found for simple potential profiles, such as the double well potential \cite{caldeirabook}, and the (biased) periodic potential \cite{caldeira2}.
These solutions hinges on the fact that the interaction of the Brownian particle couples to the thermal bath frequencies via the spectral density function
\begin{equation}
  J(\omega)
  =
  \frac{\pi}{2} \sum_{n} \frac{C_{n}^{2}}{m_{n} \omega_{n}} \delta(\omega - \omega_{n}), \quad \omega > 0.
\end{equation}
Exact solutions can be found by limiting the particle's response to the particular case of a thermal bath with ohmic profile and a high-frequency cutoff \( \Lambda > 0 \), given by
\begin{equation}
  J(\omega)
  =
  \eta \omega, \quad 0 < \omega < \Lambda.
\end{equation}
This means that for these kinds of toy models, the dissipative dynamics induced by environment interactions mostly involves the low-frequency modes of the environment.
In the Yang-Mills correspondence, that means topological information (associated with infrared degrees of freedom) %
decoheres, while localized field configurations, such as instantons, would remain intact.

%
%
In the particular case of a double-well potential \cite{caldeirabook}, for example, we can map the high-frequency dynamics of the model to an effective two-level system.
The degrees of freedom in this case are the symmetric and anti-symmetric combinations of the damped harmonic oscillator states centered at the bottom of each potential well
\begin{equation}
  \Psi_{\pm}(x, \{q_{n}\})
  =
  \frac{1}{\sqrt{2}}[\Psi_{R}(x, \{q_{n}\}) \pm \Psi_{L}(x, \{q_{n}\})].
\end{equation}
%
Generally one can separate the associated density matrix into a coherent and an incoherent part, where the coherent part can be rotated as the projection part within a certain direction $\hat{S}_z$ for the two-level system above \cite{caldeirabook}
\begin{equation}
  \hat{\rho}_{S}
  =
  \sum_{n} A_{n} \ket{n} \bra{n} + P(t) \sum_{m \neq n} B_{mn} \ket{m} \bra{n}.
\end{equation}
Since $H_I$ does not commute with $H_S$, the equation of motion for the reduced density matrix of the system becomes an initial value problem, with all time dependence can put into $P(t)$ (which could be a matrix in the ``winding number'' basis).
It can be shown then that $P(t)$ obeys the damped harmonic oscillator equation
\begin{equation}
  \label{peq}
  \ddot{P} + T^{-1} \dot{P} + \Delta^2 P =0,
\end{equation}
with $T^{-1}$ and $\Delta$ functions of the original parameters.

The exact form of the parameters can be an involved calculation, but their dependence on the length scales of the problem is universal.
As shown in \cite{caldeirabook}, while $\Delta$ is dominated by short-range physics (in our context this means instantons equation (\ref{instanton}), hence $\Delta^{-1} \sim \rho$), the damping time-scale $T$ depends on the softest scale connected to the size of the reservoir (in our context, this is the horizon radius).
The latter can be removed from the system by ``adiabatic renormalization'', where all dimensionful parameters are presented as a ratio of the cutoff frequency.
In the limit we want to reach, where the cutoff frequency dependence is very small, one needs $\alpha \rightarrow 0$ (defined in chapter 9 of \cite{caldeirabook}).
In fact, if one compares the scales in \equref{hyerarchy} to the running of \equref{peq} one sees that the applicability of the Unruh EFT in an instanton context is equivalent to the theory being near the $\alpha \rightarrow 1$ infrared fixed point.     This means in a theory having a generally covariant {infrared limit, the effective $P$ ($\sim \theta$)}  will decay to zero on a time-scale set and inversely proportional to the lowest frequency. 

To compare our models to a real quantum field theory vacuum, either Yang-Mills or its effective theory implementation, we need to be a bit more rigorous.
The path integral formalism can be connected to density matrix language via \cite{entangle}, where the similarity between QCD and the periodic potential setup is more clear.

The density matrix is related to the partition function via 
\begin{equation}
  \label{rhodef}
  \langle x| \,\rho\, |x' \rangle= \frac{1}{Z} \int \mathcal{D}\phi(t=t_0,x) <\phi|\Psi><\Psi|\phi> 
\end{equation}
\[\  = \frac{1}{Z}
  \int^{\tau=\infty}_{\tau=-\infty} \int
  \left[ \mathcal{D}\phi, \mathcal{D} y(\tau)\,\mathcal{D} y'(\tau)\right]\,e^{- iS (\phi y, y')}
\cdot \delta \left[ y(0^+) - x'\right]\,\delta \left[ y'(0^-) - x\right]\ ,\]
where $|\Psi>$ is some {eigenstate} basis and $\psi$ are field configurations to be integrated over.    

Let us now consider expand $\Psi$ in terms of $\Psi_n$, the wave functional corresponding to equation (\ref{sumnz}).
In both the QCD case and the periodic potential case,
\begin{equation}
  \label{rhopure}
  \hat{\rho} \underbrace{\rightarrow \int\mathcal{D}\phi}_{\lim_{x\rightarrow 0^+ - O^-}\mathcal{D}\phi(t=t_0,x)}   <\phi|\Psi_n><\Psi_m|\phi> \exp\left[ \theta(n-m)  \right]
\end{equation}
where $\Psi_{n}$ are Eigenstates of the Hamiltonian with, additionally, boundary condition set by the winding number (For the periodic potential, it is a fixed number of turns around $N$ wells, for the QCD case it is a given winding number).

Now, the previous section has argued that background independence implies independence from horizon terms.
In this section, the tracing over the horizon terms was argued to be equivalent to tracing over the ``infrared'' degrees of freedom $k \sim (0^+-0^-)^{-1}$ in equation (\ref{rhodef}).

Our mechanism adds to the system (S) a thermal bath (B) and in both cases
\begin{equation}
\label{rhomixed}
  \hat{\rho}_{QCD} =  \mathrm{Tr}_{B} \hat{\rho}_{S+B},
  \quad
  \hat{\rho}_{S+B} \sim c_{k k'} \left|k\right>\left<k'\right|,
  \quad
  k_{B} \sim (0^+-0^-)^{-1} \ll k_{S},
\end{equation}
where $k$ refers to momentum and $c_{kk'}$ encode all structure of the vacuum.
The arguments of the previous section make it clear that for background independence to be achieved, the tracing of the bath degrees of freedom must make no difference to the effective Lagrangian.

Let us, as above, refer to $\Psi_m$ as physical states in flat space of winding number $m$ and $\Phi_m$ as $\Psi_m$ with the infrared limit decohered (all dependence of $k_{B}$ removed, and $k_{S}$ $\sim \rho^{-1}$ of equation (\ref{instanton}) unchanged).
While $\Psi_m$ and $\Psi_n$ are not Eigenstates of the Hamiltonian, they do form an orthogonal set.
In contrast, the decoherence of $k_{bath}$ means that
\begin{equation}
  \label{overlap}
  \left<\Phi_{m}| \Phi_{n}\right> \sim Tr_k c_{k \sim k_{Bath}} \ne 0,
\end{equation}
with the fact that instanton states have diverging infrared Fourier coefficients and infinite occupation numbers ensuring a finite overlap even if the ultraviolet part of the instanton is unchanged.

Thus, the decohered density matrix $\hat{\rho}_D$, will not be diagonal in the $\left|n\right>$ basis.
Shifting them to a diagonal basis $\left|\Psi'\right>$ will involve a generally complex rotation in phase space, whose $j$th eigenvalue can be represented as complex numbers $r_j e^{i \alpha_j}$.
Putting these together we get, from equation (\ref{rhopure}), (\ref{rhomixed}) and (\ref{overlap}),
\begin{equation}
  \label{rhorot}
  \hat{\rho} \propto \sum_{n,m,j} e^{i (\theta n + \sum_j \alpha_j)}  \mathcal{D}\phi \left<\phi|\Psi_n'\right>\left<\Psi_n'|\phi\right>
\end{equation}
$\left|\Psi_n'\right>$ and $\left|\Psi_n\right>$ are distinguishable only via the IR part of the partition function and an unobservable renormalization parameter given by $r_j$.
The phase, however, was rotated by an ``infinite'' number of angles $\alpha_j$, and hence can be safely assumed to decohere.


Considering the density matrix to evolve dynamically, we would conclude the infrared form of the dynamics of this decoherence  will be controlled by a damped equation of the type equation (\ref{peq}), with the damping time of the order of the horizon parameter (the cosmological constant in a de-Sitter space, acceleration for Rindler space and so on).
This ``coincidence'' can actually be seen from the form of equation (\ref{zsourcedef}) and the argument, made in \cite{zubarev}, that general covariance of the time parameter requires correlated fluctuation (``many outcomes for an initial condition'') and dissipation (``one outcome for many initial conditions'') that transform covariantly.
Topological quantum fluctuations that fix the $\theta$ term are ``the softest scale'', set at the horizon scale.
They must be matched by an equally soft dissipation, making  such ``slow'' dynamics is unobservable.
This implies $\theta$ damping is irrelevant for any measurement $k\sim k_{Bath}$.

Finally, the interpretation of decoherence as the effect of unmeasured degrees of freedom can connect this section to the previous section \ref{gencluster}.
Because the horizon hides the information allowing the winding number to be instantaneously measured, the density matrix in the basis of $\left|n\right>$ becomes
\begin{equation}
  e^{i n \theta} \left|n\right>\left<m\right| \delta_{mn} \rightarrow \sum_{m', n'} e^{i n \theta} P(n',n) P(m',m) \left|n'\right>\left<m'\right| \delta_{m'n'},
\end{equation}
where $P(n',n)$ represent {\em semi-classical} probabilities of winding numbers disappearing behind a horizon.
Rotating bases will add complex coefficients, parametrized by the $r_j e^{i \alpha_j}$ of the previous equation equation (\ref{rhorot}).
The $r_j$'s reflect the normalization of the field strength in response to the horizon changing the configuration space.
It should be unobservable for frequencies higher than the horizon radius.
The phases $\alpha_j$ however remains and rotates $\theta$ chaotically for each momentum mode, equivalent to decoherence (see a good discussion of ``coherent'' and ``chaotic'' sources here \cite{gyul}).

In conclusion, we note that a similar mechanism to what we suggest was proposed in the 80's as an origin of {\em spontaneously broken} symmetries \cite{broksym}.
The success of the Higgs model and chiral symmetry breaking invalidated further development of this idea, but for {\em anomalously broken} symmetries it has potential.

\section{Discussion}

The previous section \ref{caldeira} makes it clear that any tracing over of degrees of freedom outside a causal horizon generally results in the asymptotic relaxation of a density matrix of a theory with connected topological sectors to a density matrix where they are connected incoherently.
This means that equation (\ref{qcdvac0}) is therefore updated to
\begin{equation}
  \label{qcdvac}
  \sum_{m,n} e^{i (m - n) \theta} \ket{m} \bra{n}
  \rightarrow
  \sum_{n} c_{n} \ket{n} \bra{n}
\end{equation}
where $ c_{n}$ doesn't depend on $\theta$ and is put to unity via field strength renormalization.  $\theta=0$ in this regard is ``special'' because coherent and incoherent summation is indistinguishable.

Note that instantons, as any ``UV'' degrees of freedom, are preserved (as they should be if there is any hope of the idea presented in this paper to be correct, since they are necessary for phenomenology (the $\eta'$ mass issue) \cite{shuryak,hooft,witten1979}, seen on the lattice \cite{lattice} and hinted at in experiment \cite{instexp,cme}).

This can be seen as a consequence of local Lorentz invariance, since, unlike the local Euclidean signature (where \equref{euclmink} holds), the infrared limit is determined by the spacetime global horizon structure as well as the local quantum field theory correlations.   Thus, \equref{qcdvac} only modifies the infrared sector of the theory, where the infrared scale could be affected by the curvature as per \equref{hyerarchy}, while local instanton fluctuations and topological susceptibility should therefore not be different from analytical continuations from Euclidean space used in \cite{shuryak,lattice,instexp,cme}.
This also means that theories where CP violation is due to a condensate, such as the electroweak sector of the standard model, do not suffer from the ambiguities discussed in this paper, since there the hyerarchy defined in \equref{hyerarchy} is well defined, with the equivalent of $\mu$ in \equref{hyerarchy} being absent and the condensate gap energy taking the place of $\rho^{-1}$. In that situation, to check general covariance one just has to take interactions with the condensate into account properly in both the inertial and the co-moving frame, a non-trivial procedure discussed in \cite{neutcp,neutus}.

If a universe is ``prepared'' with a finite $\theta$ and a horizon, the timescale for the $\theta$ to relax to its effectively zero value can be given by the methods of section \ref{caldeira}.
The asymptotic density matrix picture is also covariant under general coordinate transformations, at least for non-inertial transformations having an acceleration smaller than the horizon size, according to section \ref{gencluster}.

The mechanism described here could be valid, however, even if general covariance is broken by quantum effects.
For instance, it could arise dynamically in a cosmological scenario.
In a ``long'' high cosmological constant phase in the early universe, natural in the slow-roll inflation scenario \cite{weinbergcosmo}, the topological sector of the QCD vacuum would have time to decohere if this phase is also deconfined (either due to temperature of a dS constant above $T_c$).
Afterwards, the decohered $\theta=0$ state would be ``locked'', since long wavelength colored perturbations which cause tunneling would be below the confinement mass gap.
Such a dynamical non-perturbative QFT problem is of course well outside this work's scope, but qualitatively this scenario might be implementable.

An obvious drawback of the explanation presented here is its lack of falsifiability.
Unlike with more traditional mechanisms of the resolution of the $\theta$ problem, we do not predict new particles, and is founded on a fundamental modification of the quantum field theory vacuum state \cite{gtholo} that in turn generally does not produce verifiable predictions.
However, one might be able to experimentally test our model with analogue systems.
Provided a fluid with internal symmetries exhibiting topological properties similar to Yang-Mills theory is found (for example a fluid with polarization \cite{spinhydro}), one might be able to put this fluid in Minkowski and de-Sitter configurations \cite{mosna}.
An effective $\theta$ term could then appear in the former and disappear in the latter, decohered by Hawking sound.  The violation of unitarity that we believe is implicit in the structure of spacetime would here arise out of the fluid dynamics limit.
In this regard,the impossibility of relativistic local equilibrium in ``polymeric fluids'' where spin and vorticity are not parallel at thermal equilibrium, pointed out in \cite{spinhydro} is a useful analogy, as in this system local equilibrium is impossible because of topologically constrained local Goldstone modes.  If the reasoning in this paper is broadly correct, we would predict that no topological insulators are possible which are also perfect liquids.  As far as we know this is indeed the case to date.

Since the effect suggested here involves integrating out infrared degrees of freedom, there might be a parallel to non-perturbative renormalization group approaches, investigated elsewhere \cite{rggroup} and motivated by the analogy between renormalization group scheme independence and general covariance \cite{polch,schafer}.  In both cases, the approach does not rely on new physics, but rather on making sure the underlying theory's local physics is insensitive to unobserved infrared perturbations.

In conclusion, we have heuristically discussed the topology of a quantum field theory in curved space.
We have argued that the presence of causal horizons, or, equivalently, the expectation of general covariance of the theory at the quantum level, generally require the asymptotic wave functional of the theory to have different topological sectors incoherently summed.
This is just what one needs to require the $\theta$ angle of the theory to vanish, in accordance with observations.
While we are very far from proving that this is the correct explanation of the strong CP problem, it is a suggestion that merits further development.

GT acknowledges support from FAPESP proc.\ 2017/06508-7, participation in FAPESP tematico 2017/05685-2 and CNPQ bolsa de produtividade 301432/2017-1.
HT Would like to acknowledge the support from the Conselho Nacional de Desenvolvimento Científico e Tecnológico (CNPq), proc. 141024/2017-8.
\appendix
\section{Toy models}
In this section we shall describe toy models which include a boundary term and the possibility to calculate the density matrix and the various forms of entropy explicitly.   They are limited in that they do not allow us to fully understand how topological terms are generated by the dynamics, since for such 1+1 theories topology arises out of boundary conditions rather than dynamics, and is decoupled from the bulk.

However,  dynamical coupling between bulk and boundary can be put in by hand in an adiabatic way which generates no entropy.
In this case, it becomes clear that the ``geometric'' and ``statistical'' definitions of entropy do not match, thereby giving an example of the tension between general covariance and topological terms.

\subsection{Quantum particle on a ring with a magnetic field \label{ring}}


We will base {ourselves} on \cite{laflamme} and Gregory Moore's lectures on Chen-Simons theory \cite{moore}.

We consider the quantum mechanical system of a particle in a ring \( S^{1} \), characterized by the Euclidean action
\begin{equation}
  \label{eq:PiaR-action}
  i S
  =
  \int dt \left(
    \frac{1}{2} I \dot{\phi}^{2}
    + \bfield \dot{\phi}
  \right)
  \longrightarrow
  - S_{E}
  =
  - \int d\tau \left(
    \frac{1}{2} I \dot{\phi}^{2}
    - i \bfield \dot{\phi}
  \right)
\end{equation}
for the angular variable \( \phi \).
The periodic-boundary conditions characterizing the system are given by \( \phi(\tau + \beta) = \phi(\tau) \) and \( \phi \sim \phi + 2\pi n \).
This theory then represents a \( (0 + 1) \) dimensional field theory \( \phi : S^{1} \rightarrow S^{1} \), which is a topological theory since \( \pi_{1}(S^{1}) = \mathbb{Z} \).
The topological charge is represented by how many times around the circle the particle runs on the course of one period \( \beta \) -- counted by the \( i \bfield \dot{\phi} \) term on the action.

We use this model as a toy example for the Laflamme procedure \cite{laflamme} of obtaining a description for the density matrix of a (free) field theory in an accelerated frame. 
This procedure is described as follows:
The comoving-time of an accelerated observer is periodic in Euclidean signature, due to the natural invariance of Minkowski spacetime under boosts with imaginary-boost parameter.
That means the accelerated observer's proper-time is represented in Euclidean signature by the angular variable in a polar-coordinate representation of spacetime.
The presence of causal horizons for the accelerated observer is reflected on the fact that you can't cover the circle with a single coordinate chart:
That can be seen from the transformation \( \tau \rightarrow \tau + \pi \) on the circle, which maps the points \( (\tau + \pi, r) \rightarrow (\tau, -r) \).
Under reverse Wick rotation \( \tau \rightarrow i \tau \), that transformation maps a Rindler wedge into {its} causal complement (that means mapping the right wedge into left wedge and vice versa).

\begin{figure}
  \centering
  \includegraphics[width=5cm]{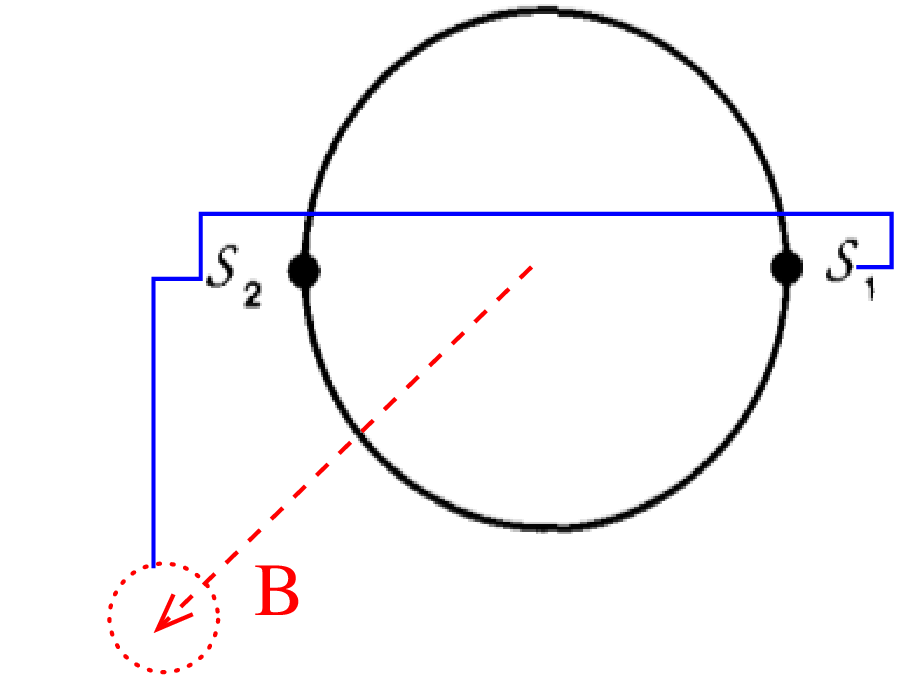}
  \caption{\small Figure from \cite{laflamme}, representing the Euclidean section of spacetime representing an accelerated observer. \( \mathcal{M}_{\pm} \) are the pieces representing the right and left wedges respectively, and \( \mathcal{S}_{k} \) the boundaries associated with the presence of causal horizons. The circumference of the circle is \( \beta \), and \( \tau_{\mathcal{S}_{2}} = \tau_{\mathcal{S}_{1}} + \beta / 2 \).}  We also include the magnetic field term $\bfield$ and the boundary regulator described at the end of this section, leading to \equref{gens}.  These are represented by the circuit connecting $\bfield$ to $S_{1,2}$
  \label{figcircle}
\end{figure}

Now we can describe the causally disconnected regions of Minkowski space by considering a division of the circle into two sections, with the common boundary (shown in figure (\ref{figcircle})) representing the causal horizon structure of the accelerated frame.
To obtain a description of physics from the point of view of the accelerated observer, we integrate out the contributions from the inaccessible region.

In mathematical terms, that corresponds to describing the theory on \( \mathcal{M}_{+} \) by the path-integral summing over the complementary region \( \mathcal{M}_{-} \):
Let \( \Psi_{\pm}(\tilde{\phi}_{1}, \tilde{\phi}_{2}) \) be the partition function for the fields on \( \mathcal{M}_{\pm} \) satisfying the boundary conditions \( \phi(\mathcal{S}_{1}) = \tilde{\phi}_{1} \) and \( \phi(\mathcal{S}_{2}) = \tilde{\phi}_{2} \).
The path integral we are interested in evaluating is, then, given by
\begin{equation}
  \label{eq:def-density-matrix}
  \rho(\tilde{\phi}_{1}, \tilde{\phi}_{1}^{\prime})
  =
  \int \mathcal{D} \tilde{\phi}_{2}
  \Psi_{+}(\tilde{\phi}_{1}, \tilde{\phi}_{2})
  \Psi_{-}(\tilde{\phi}_{1}^{\prime}, \tilde{\phi}_{2}),
\end{equation}
where \( \rho \) is the density matrix associated with the states \( \tilde{\phi}_{1}, \tilde{\phi}_{1}^{\prime} \in \mathcal{M}_{+} \) accessible to the observer.

In the context of quantum field theory, figure (\ref{figcircle}) is only a representation of the true Euclidean section of spacetime, whose spatial dimensions compose the correct picture in higher dimensions.
However, for free theories such as the scalar field
\begin{equation}
  S
  =
  \frac{1}{2} \int d^{D + 1}x \sqrt{g} \left(
    g^{\mu \nu} \nabla_{\mu} \Phi \nabla_{\nu} \Phi
    + m^{2} \Phi^{2}
  \right),
\end{equation}
we can make use of the Fourier transform and study a single mode of the field \cite{laflamme}
\begin{equation}
  S_{E}^{\lambda}
  =
  \frac{1}{2} \int d\tau \left(
    \dot{\phi}_{\lambda}^{2}
    + \lambda \phi_{\lambda}^{2}
  \right),
\end{equation}
which reduces to a quantum harmonic oscillator in a ring.
Given the close connection with Euclidean-signature quantum mechanics on a ring with the physics of accelerated observers, we use this analogy to study the topological system from equation (\ref{eq:PiaR-action}) in an accelerated frame.

Our target is therefore calculating the density matrix in equation (\ref{eq:def-density-matrix}) for the model described by the action (\ref{eq:PiaR-action}) for the particle in the ring.
First, then, we calculate the partition functions
\begin{equation}
  \Psi_{\pm}(\tilde{\phi}_{1}, \tilde{\phi}_{2})
  =
  \int_{\mathcal{C}[\mathcal{M}_{\pm}]} \mathcal{D}\phi \,
  e^{- S_{E}}
\end{equation}
for the \( (0+1) \)-dimensional field \( \phi \) with corresponding boundary conditions
\begin{equation}
  \label{eq:piar-boundary-classes}
  \mathcal{C}[\mathcal{M}_{\pm}]
  \equiv
  \{
    \phi(\mathcal{S}_{1}) = \tilde{\phi}_{1}
    \text{ and }
    \phi(\mathcal{S}_{2}) = \tilde{\phi}_{2}
  \}.
\end{equation}
Being a free theory, the path-integral can be solved by means of the solutions to the Euclidean equations of motion
\begin{equation}
  \label{eq:piar-euclidean-eom}
  \frac{d^{2}}{d \tau^{2}} \phi_{cl}(\tau) = 0
\end{equation}
that extremizes the action.
By virtue of the topological nature of the theory, the solutions of this classical equation of motion are parametrized by an integer \( n \in \mathbb{Z} \) that counts how many times the particle winds around the circle for given boundary conditions.
Let \( \phi_{cl}^{\pm} \) be the classical solutions to the Euclidean equations of motion (\ref{eq:piar-euclidean-eom})%
%
\begin{equation}
\begin{split}
  \phi_{cl}^{+}(\tau)
  &=
  \tilde{\phi}_{1}
  + \frac{2}{\beta} \left(
    \tilde{\phi}_{2}
    - \tilde{\phi}_{1}
    + 2 \pi n_{+}
  \right) \tau
  \\
  \phi_{cl}^{-}(\tau)
  &=
  2 \tilde{\phi}_{2} - \tilde{\phi}_{1} - 2\pi n_{-}
  + \frac{2}{\beta} \left(
    \tilde{\phi}_{1}
    - \tilde{\phi}_{2}
    + 2 \pi n_{-}
  \right) \tau
\end{split}
\end{equation}
satisfying the boundary conditions (\ref{eq:piar-boundary-classes}).
The integers \( n_{\pm} \in \mathbb{Z} \) are the winding numbers of the field configurations, which appear in the combination \( 2\pi n_{\pm} \) since \( \phi(\tau) \) lives on the circle \( S^{1} \).
The field configurations
\begin{equation}
  \exp \left\{i \phi_{cl}^{\pm}(\tau)\right\} \in S^{1}
\end{equation}
are called instantons, and they interpolate between the classical, \( \tau \)-independent, zero-action ``vacuum'' solutions \( \tilde{\phi}_{cl}^{vac}(\tau) = \tilde{\phi}_{1/2} \), on each \( \mathcal{M}_{\pm} \) manifold.

The angular velocity of the classical solutions take the form
\begin{equation}
  \dot{\phi}_{cl}^{\pm}(\tau)
  =
  \frac{4 \pi}{\beta} \left(
    n_{\pm} \pm \frac{\tilde{\phi}_{2}
    - \tilde{\phi}_{1}}{2\pi}
  \right),
\end{equation}
and the classical action for each instanton solution is given by the expression
\begin{equation}
\begin{split}
  \label{eq:piar-instanton-action}
  - S_{E}^{\pm}
  &=
  - \int_{\mathcal{C}[\mathcal{M}_{\pm}]} d\tau \left\{
    \frac{1}{2} I \left(\frac{4\pi}{\beta}\right)^{2} \left(
      n_{\pm}
      \pm \frac{\tilde{\phi}_{2}
      - \tilde{\phi}_{1}}{2\pi}
    \right)^{2}
    - \frac{4 \pi i \bfield}{\beta} \left(
      n_{\pm}
      \pm \frac{\tilde{\phi}_{2} - \tilde{\phi}_{1}}{2\pi}
    \right)
  \right\}
  \\
  &=
  - \frac{4 \pi^{2} I}{\beta} \left(
    n_{\pm}
    \pm \frac{\tilde{\phi}_{2} - \tilde{\phi}_{1}}{2\pi}
  \right)^{2}
  + 2 \pi i \bfield \left(
    n_{\pm}
    \pm \frac{\tilde{\phi}_{2} - \tilde{\phi}_{1}}{2\pi}
  \right).
\end{split}
\end{equation}
%

Now, the partition functions \( \Psi_{\pm}(\tilde{\phi}_{1}, \tilde{\phi}_{2}) \) are given by the sum over all possible field configurations \( \phi^{\pm} \), and all such field configurations falls under a particular equivalence class determined by its winding number \( n[\phi^{\pm}] \in \mathbb{Z} \).
That means we must sum over all classical instanton solutions, taken to be the representatives of each winding-sector, as well as all possible fluctuations
\begin{equation}
  \delta \phi^{\pm} = \phi - \phi_{cl}^{\pm}
\end{equation}
around these classical backgrounds.
The fluctuating fields \( \delta \phi^{\pm} \) are topologically trivial, meaning \( n[\delta \phi^{\pm}] = 0 \), and satisfy the boundary conditions
\begin{equation}
  \delta \phi^{\pm}(\mathcal{S}_{i}) = 0.
\end{equation}
Since we are expanding around the solutions to the classical equations of motion, the total action is merely the sum
\begin{equation}
  S[\phi] = S[\phi_{cl}^{\pm}] + S[\delta \phi^{\pm}],
\end{equation}
and therefore we obtain the factorization
\begin{equation}
  \label{eq:piar-partition-func-factorization}
  \Psi_{\pm}(\tilde{\phi}_{1}, \tilde{\phi}_{2})
  =
  \int {\mathcal{D} \phi}\,
  e^{- S_{E}}
  \sim
  Z_{\delta \phi^{\pm}}
  \sum_{n_{\pm} \in \mathbb{Z}}
  e^{- S_{E}[\phi_{cl}^{\pm}]},
\end{equation}
with \( Z_{\delta \phi^{\pm}} \) the (topologically trivial) partition function for the fluctuating field \( \delta \phi^{\pm} \)
\begin{equation}
  Z_{\delta \phi^{\pm}}
  =
  \int_{\delta \phi^{\pm}(\mathcal{S}_{1}) = 0}^{\delta \phi^{\pm}(\mathcal{S}_{2}) = 0}
  \mathcal{D} \delta \phi^{\pm}
  e^{- S[\delta \phi^{\pm}]}.
\end{equation}

Since we are interested in the topological properties of the reduced density matrix, and \( Z_{\delta \phi^{\pm}} \) is only part of the normalization of \( \Psi_{\pm} \), we only deal with the instanton sum of equation (\ref{eq:piar-partition-func-factorization}).
Substituting the expression for the classical action of each instanton solution (\ref{eq:piar-instanton-action}), we get
\begin{equation}
  \label{eq:piar-partition-func-instantons}
  \Psi_{\pm}(\tilde{\phi}_{1}, \tilde{\phi}_{2})
  \sim
  \sum_{n_{\pm}}
  \exp \left\{
    - \frac{4 \pi^{2} I}{\beta} \left(
      n_{\pm}
      \pm \frac{\tilde{\phi}_{2} - \tilde{\phi}_{1}}{2\pi}
    \right)^{2}
    + 2 \pi i \bfield \left(
      n_{\pm}
      \pm \frac{\tilde{\phi}_{2} - \tilde{\phi}_{1}}{2\pi}
    \right)
  \right\}.
\end{equation}

Now we are in a position to evaluate the reduced density matrix (\ref{eq:def-density-matrix}) obtained by integrating out all \( \mathcal{M}_{-} \) fields.
An elegant  approach is to make use of the representation of the partition functions \( \Psi_{\pm} \) in terms of the Riemann's theta function 
\begin{equation}
  \mathcal{R}
  \begin{bmatrix}
    n_{0} \\
    z_{0}
  \end{bmatrix}	
  (z, \tau)
  =
  \sum_{n \in \mathbb{Z}}
  \exp \left(
    i \pi \tau (n + n_{0})^{2}
    + 2\pi i (n + n_{0}) (z + z_{0})
  \right).
\end{equation}
Writing \( \Delta \tilde{\omega} = (\phi_{2} - \phi_{1}) / 2\pi \), we get the compact expression for the wedge partition functions
\begin{equation}
  \Psi_{\pm}(\tilde{\phi}_{1}, \tilde{\phi}_{2})
  \sim
  \mathcal{R}
  \begin{bmatrix}
    \pm \Delta \tilde{\omega} \\
    \bfield
  \end{bmatrix}	
  \left(0, \frac{4\pi i I}{\beta}\right).
\end{equation}
Using the transformation property of the \( \mathcal{R} \) function under modular transformations
\begin{equation}
  \mathcal{R}
  \begin{bmatrix}
    n_{0} \\
    z_{0}
  \end{bmatrix}	
  \left(z, \tau\right)
  =
  (- i \tau)^{- 1/2}
  e^{2\pi i z_{0} n_{0}}
  e^{- i \pi z^{2} / \tau}
  \mathcal{R}
  \begin{bmatrix}
    z_{0} \\
    - n_{0}
  \end{bmatrix}	
  \left( \frac{- z}{\tau}, \frac{- 1}{\tau} \right),
\end{equation}
we obtain an alternative and equivalent representation
\begin{equation}
  \mathcal{R}
  \begin{bmatrix}
    \pm \Delta \tilde{\omega} \\
    \bfield
  \end{bmatrix}	
  \left(0, \frac{4\pi i I}{\beta}\right)
  =
  \left(\frac{4\pi I}{\beta}\right)^{- 1/2}
  e^{\mp 2\pi i \bfield \Delta \tilde{\omega}}
  \mathcal{R}
  \begin{bmatrix}
    \bfield \\
    \mp \Delta \tilde{\omega}
  \end{bmatrix}	
  \left(0, - \frac{\beta}{4\pi i I}\right).
\end{equation}
Expanding this expression in terms of the sum representation of \( \mathcal{R} \) we end up with the relationship
\begin{multline}
  \Psi_{\pm}(\tilde{\phi}_{1}, \tilde{\phi}_{2})
  \sim
  \sum_{n_{\pm}}
  \exp \left\{
    - \frac{4 \pi^{2} I}{\beta}
    \left( n_{\pm} \pm \Delta \tilde{\omega} \right)^{2}
    + 2 \pi i \bfield \left(
      n_{\pm}
      \pm \Delta \tilde{\omega}
    \right)
  \right\}
  = \\ =
  \left(\frac{4\pi I}{\beta}\right)^{- 1/2}
  \sum_{n_{\pm}}
  \exp \left\{
    - \frac{\beta}{4I} (n_{\pm} - \bfield)^{2}
    \pm 2\pi i n_{\pm} \Delta \tilde{\omega}
  \right\}.
\end{multline}

Now we can proceed to integrate out \( \tilde{\phi}_{2} \) using the new representation of \( \Psi_{\pm} \).
In this representation, the reduced density matrix is given by the expression
\begin{multline}
  \rho(\tilde{\phi}_{1}, \tilde{\phi}_{1}^{\prime})
  \sim
  \int_{0}^{2\pi} d \tilde{\phi}_{2}
  \Psi_{+}(\tilde{\phi}_{1}, \tilde{\phi}_{2})
  \Psi_{-}(\tilde{\phi}_{1}^{\prime}, \tilde{\phi}_{2})
  \sim \\ \sim
  \left(\frac{\beta}{4\pi I}\right)
  \int_{0}^{2\pi} {d \tilde{\phi}_{2}}\,
  \sum_{n_{\pm} \in \mathbb{Z}}
  \exp \left\{
    - \frac{\beta}{4 I}
    \left( n_{+} - \bfield \right)^{2}
    + i n_{+} \left(\tilde{\phi}_{2} - \tilde{\phi}_{1} \right)
  \right\}
  \times \\ \times
  \exp \left\{
    - \frac{\beta}{4 I}
    \left( n_{-} - \bfield \right)^{2}
    - i n_{-} \left(\tilde{\phi}_{2} - \tilde{\phi}_{1}^{\prime} \right)
  \right\},
\end{multline}
and by using the integral representation of the Kronecker delta
\begin{equation}
  \delta_{mn}
  =
  \frac{1}{2\pi}
  \int_{0}^{2\pi} d\tilde{\phi}_{2}\,
  e^{i (m - n) \tilde{\phi}_{2}}
\end{equation}
we obtain
\begin{equation}
  \rho(\tilde{\phi}_{1}, \tilde{\phi}_{1}^{\prime})
  \sim
  \left(\frac{\beta}{2 I}\right)
  \sum_{n \in \mathbb{Z}}
  \exp \left\{
    - \frac{\beta}{2 I}
    \left(n - \bfield \right)^{2}
    - i n \left(\tilde{\phi}_{1}  - \tilde{\phi}_{1}^{\prime} \right)
  \right\}.
\end{equation}
Once again we normalizing the density matrix by demanding unity trace
\begin{equation}
  \Tr \rho
  =
  \int_{0}^{2\pi} d \tilde{\phi} \,
  \rho(\tilde{\phi}, \tilde{\phi})
  =
  1,
\end{equation}
and that yields the expression for the matrix elements of the density matrix
\begin{equation}
  \label{eq:piar-final-density-matrix-riemann}
  \rho(\tilde{\phi}_{1}, \tilde{\phi}_{1}^{\prime})
  =
  \frac{1}{2\pi Z}
  \sum_{n \in \mathbb{Z}}
  \exp \left\{
    - \frac{\beta}{2 I}
    \left(n - \bfield \right)^{2}
    - i n \left(\tilde{\phi}_{1}  - \tilde{\phi}_{1}^{\prime} \right)
  \right\},
\end{equation}
\begin{equation}
  Z
  =
  \sum_{n \in \mathbb{Z}}
  \exp \left\{
    - \frac{\beta}{2 I}
    \left(n - \bfield \right)^{2}
  \right\}.
\end{equation}
%


We could have obtained this result more directly from the expression of the Euclidean propagator
\begin{equation}
  \langle \tilde{\phi}_{1}^{\prime} |
  e^{- \beta H}
  | \tilde{\phi}_{1} \rangle
  =
  \frac{1}{2\pi} \sum_{n \in \mathbb{Z}}
  \exp \left\{
    - \frac{\beta}{2I} (n - \bfield)^{2}
    - i n (\tilde{\phi}_{1} - \tilde{\phi}_{1}^{\prime})
  \right\}.
\end{equation}
This propagator can be obtained from the system Hamiltonian
\begin{equation}
  H
  =
  \frac{1}{2I}
  \left(- i \partial_{\phi} - \bfield\right)^{2}
\end{equation}
whose spectrum consists in the eigenvectors \( \{\ket{m}, m \in \mathbb{Z}\} \), given by
\begin{equation}
  \langle \phi | m \rangle
  =
  \Phi_{m}(\phi)
  =
  \frac{1}{\sqrt{2\pi}} e^{i m \phi}.
\end{equation}
The procedure of integrating out half of the spacetime circle corresponds to the process of separating \( \beta = \frac{\beta}{2} + \frac{\beta}{2} \) and introducing a complete set of states in between
\begin{multline}
  \langle \tilde{\phi}_{1}^{\prime} |
  e^{- \beta H}
  | \tilde{\phi}_{1} \rangle
  =
  \int_{0}^{2\pi} d \tilde{\phi}_{2} \,
  \langle \tilde{\phi}_{1}^{\prime} |
  e^{- \frac{1}{2} \beta H}
  | \tilde{\phi}_{2} \rangle
  \langle \tilde{\phi}_{2} |
  e^{- \frac{1}{2} \beta H}
  | \tilde{\phi}_{1} \rangle
  \\ =
  \int_{0}^{2\pi} d \tilde{\phi}_{2} \,
  \Psi_{+}(\tilde{\phi}_{1}, \tilde{\phi}_{2})
  \Psi_{-}(\tilde{\phi}_{1}^{\prime}, \tilde{\phi}_{2})
  \sim
  \rho(\tilde{\phi}_{1}, \tilde{\phi}_{1}^{\prime}).
\end{multline}

Following this, we proceed to calculate thermodynamical quantities from the model.
First, we note that we can write the identity in two equivalent ways
\begin{equation}
  \int_{0}^{2\pi} d\phi | \phi \rangle \langle \phi |
  =
  \sum_{m \in \mathbb{Z}} | m \rangle \langle m |
  =
  \mathbb{I}.
\end{equation}

For convenience, we note that (\ref{eq:piar-final-density-matrix-riemann}) can be rewritten in terms of the eigenfunctions \( \Phi_{m}(\phi) \) of the Hamiltonian
\begin{equation}
\begin{split}
  \langle \phi^{\prime} | \rho | \phi \rangle
  = 
  \frac{1}{2\pi Z}
  \sum_{n \in \mathbb{Z}}
  \exp \left\{
    - \frac{\beta}{2 I}
    \left(n - \bfield \right)^{2}
    - i n \left(\phi  - \phi^{\prime} \right)
  \right\} 
  = \frac{1}{Z}
  \sum_{n \in \mathbb{Z}}
  e^{- \beta E_{n}}
  \Phi_{n}(\phi)
  \Phi_{n}^{*}(\phi^{\prime}).
\end{split}
\end{equation}
We can also find the matrix elements of the density matrix in the Hamiltonian eigenstate basis
\begin{equation}
\begin{split}
  \langle m | \rho | n \rangle
=
  \frac{1}{Z} e^{- \beta E_{m}} \delta_{mn}.
\end{split}
\end{equation}
Note that in the basis of Hamiltonian eigenstates \( \rho \) is diagonal, and we find that
\begin{equation}
  \rho = \frac{1}{Z} e^{- \beta H},
\end{equation}
which indeed confirms that \( \rho \) represents a finite temperature system of excitations of \( H \).
The energy and entropy of the system are obtained by using the diagonal representation of \( \rho \) in the basis of eigenstates of \( H \), and are given by
%
\begin{equation}
  E
  \equiv
  \langle H \rangle
  \equiv
  \Tr \rho H
  =
  \frac{1}{Z} \sum_{m \in \mathbb{Z}}
  E_{m} e^{- \beta E_{m}}
\end{equation}
in this picture we can calculate the Von Neumann entropy and recover the usual thermodynamic relation
\begin{equation}
\begin{split}
  S
  \equiv
  - \langle \ln \rho \rangle
  \equiv
  - \Tr \rho \ln \rho =
  \beta \langle H \rangle + \ln Z.
\end{split}
\end{equation}
Differentiating the entropy with respect to the internal energy we get the expected relation
\begin{equation}
  \label{eq:piar-dS/dE}
  \frac{\partial S}{\partial E}
  =
  \beta,
\end{equation}
Confirming that indeed \( \beta^{-1} = T \) represents the temperature of the system.
{Taking the entropy to be an explicit function of both \( \beta \) and \( \bfield \) we obtain the usual thermodynamical expression
\begin{equation}
  dS
  =
  \left(\frac{\partial S}{\partial \beta}\right)_{\!\!\bfield} d\beta
  + \left(\frac{dS}{d \bfield}\right)_{\!\!\beta} d \bfield=
  \beta\, dE
  + \left(\frac{\partial S}{\partial \bfield}\right)_{\!\!E} d \bfield.
\end{equation}
}
Observing equations (\ref{eq:piar-final-density-matrix-riemann}), it would appear that that the \( \bfield \)-term on the density matrix doesn't depend on the temperature of the field \( \beta \sim T^{-1} \).
That means the topological sector, in this particular case, does not seem to be sensitive to the Hawking-Unruh temperature associated with the Euclidean-periodicity in proper-time.
In this toy-model, the magnetic field \( \bfield \) takes the role of the theta-parameter in other scenarios, such as Chern-Simons theory and the CP-violating \( \theta QCD_{4} \) scenario.

On the other hand, 
in real 3+1 field theory $\bfield$ is determined dynamically from the theory.   While we cannot reproduce this effect explicitly, we can ``mock it up'' and study its consequences in the following way:  let us hook up a system prepared as in Fig. \ref{figcircle}, a quantum particle on a circle in equilibrium, to an adiabatic device measuring heat capacity.  This device in turn regulates a magnetic solenoid generating the field $\bfield$.  This device is adiabatic and does not measure any microstates, so, unlike ``Maxwell's demon type setups'', its added entropy content is negligible.

What it does, however, is bring the system away from the thermodynamic limit, so the thermodynamic entropy is not automatically the Legendre transform of the energy.  Instead, it is defined by
\begin{equation}
\beta' = \left. \frac{d S}{dE}\right|_{\hat{n}}
\end{equation}
where $\hat{n}$ is the direction in $\bfield,E$ space where two systems hooked up in parallel stop.

{Considering the tangent vector \( \hat{n} \) as the derivative of a curve \( \gamma(\lambda) = (E(\lambda), \mathcal{B}(\lambda)) \) at \( \lambda = 0 \), we get \( \beta^{\prime} = \gamma^{\prime}(0) \cdot \nabla S \).
  Parameterizing the curve with the internal energy we get that the equilibrium configuration of such a system is related to the usual equilibrium as
%
\begin{equation}
  \label{gens}
  \beta^{\prime}(E, \mathcal{B})
  \equiv
  \hat{n} \cdot \nabla S
  =
  \beta
  + \left(\frac{\partial S}{\partial \mathcal{B}}\right)_{\!\!E} \frac{d \mathcal{B}}{dE}.
\end{equation}
}
The moral of the story here is that if there is coupling between bulk and topological degrees of freedom, one could expect that geometric and thermodynamic temperatures are different.    While this was a quantum problem without a ``real'' Unruh effect, in the next section we shall generalize this reasoning to a one dimensional quantum field theory.
\subsection{Maxwell theory in (1+1) dimensions \label{maxwell}}
Let us now consider the 1+1D Maxwell theory examined in \cite{adam,blomm,blomm2}.
This is a quantum field theory, where the Unruh effect is possible.
At first site, the equivalence of Eq. 2.28 of \cite{blomm} for the topological configuration with Eq. 49 of \cite{calixto}, together with the discussion following Eq. 49, would seal our case that a non-trivial topological charge will spoil the Unruh effect.  However, this is not the case, since the theory examined in \cite{blomm} has no propagating degrees of freedom.

In fact, this theory can be mapped on the problem examined in the previous section, since pure electrodynamics in \( (1 + 1) \) dimensions is a topological theory, since \( \pi_{1}(U(1)) = \mathbb{Z} \).
This means pure electrodynamics admits a topological $\theta$ term, similar in nature to the QCD case, given by
\begin{equation}
  \mathcal{L}
  =
  - \frac{1}{4} F_{\mu \nu} F^{\mu \nu}
  - \frac{e \theta}{4\pi} \epsilon_{\mu \nu} F^{\mu \nu}
  \equiv
  \frac{1}{2} \efield^{2} + \frac{e \theta}{2\pi} \efield,
  \label{lblom}
\end{equation}
where \( \efield \equiv F_{01} \).
Since the Lagrangian only depends on \( \efield \), we can treat it like the canonical field and find its equation of motion
\begin{equation}
  \efield = - \frac{e \theta}{2\pi}.
\end{equation}
Since the field is just a constant \( \mathbb{C} \)-number, the theory is trivial.
That means there is only one physical state, the vacuum state \( \ket{\theta} \), satisfying
\begin{equation}
  \efield \ket{\theta}
  =
  - \frac{e \theta}{2\pi} \ket{\theta},
  \quad
  \mathcal{H} \ket{\theta}
  =
  \frac{1}{2} \left(\frac{e \theta}{2\pi}\right)^{2} \ket{\theta} \eqcomma \dot{F}=0
  \label{estatic}
\end{equation}
In other words, \( \theta \) is the simultaneous eigenstate of the Hamiltonian density and the electric field, and $\theta$ can be interpreted as a constant background electric field \( \mathcal{E} = \frac{e \theta}{2\pi} \).
In terms of \( \mathcal{E} \), that means the Hamiltonian density takes the form
\begin{equation}
  \mathcal{H} \ket{\mathcal{E}}
  =
  \frac{\mathcal{E}^{2}}{2} \ket{\mathcal{E}}.
\end{equation}
Furthermore, as shown in \cite{blomm2}, this theory on a disk can be  mapped exactly to the particle on a ring problem of the previous section.
From the electromagnetic duality in \( (1 + 1) \) dimensions, the Euclidean partition function of the theory in Eq.\ (\ref{lblom}) is related to the particle in a ring by
\begin{equation}
\begin{split}
  Z
  &=
  \int \mathcal{D} A \exp \left\{
    - \int d^{2} x \left(
      \frac{1}{4} F_{\mu \nu} F_{\mu \nu} - i \frac{e \theta}{2\pi} \epsilon_{\mu \nu} F_{\mu \nu}
    \right)
  \right\} \\
  &=
  \int \mathcal{D} \mathcal{E} \delta(\dot{\mathcal{E}})
  \exp \left\{
    - \int_{0}^{\beta} d\tau\, \left(
      \frac{a}{2} \mathcal{E}^{2} - i \frac{e \theta}{2\pi} \mathcal{E}
    \right)
  \right\} \\
  &\equiv
  \int \mathcal{D} \phi
  \exp \left\{
    - \int_{0}^{\beta} d\tau \left(
      \frac{1}{2a} \dot{\phi}^{2}
      - i \frac{e \theta}{2\pi} \dot{\phi}
    \right)
  \right\},
\end{split}
\end{equation}
where \( a \) is defined in relation to the area enclosed by the ring \( a = \beta^{-1} \int \sqrt{g} \).
Thus, the correspondence becomes
\begin{equation}
  \mathcal{E} \rightarrow \dot{\phi},
  \quad
  a \rightarrow I^{-1},
  \quad \text{and} \quad
  \frac{e \theta}{2\pi} \rightarrow \mathcal{B}.
\end{equation}
%
As seen in \cite{entangle} the partition function requires asymptotic states as well as the action.  However, we can use this equivalence as a starting point to calculate exactly the vacua in inequivalent frames.

Let us consider such a field prepared to be in a Minkowski vacuum state.
By means of equation (4.24) of \cite{blomm}, defining the action of the Wilson line on the wedge-decomposition of a state as
\begin{equation}
  \mathcal{W}_{e} \ket{\Psi}
  =
  \mathcal{W}_{e} \frac{1}{Z} \sum_{\mathcal{E}} e^{- A \frac{\mathcal{E}^{2}}{2}} \ket{\mathcal{E}}_{L} \otimes \ket{\mathcal{E}}_{R}
  =
  \frac{1}{Z} \sum_{\mathcal{E}} e^{- A \frac{\mathcal{E}^{2}}{2}} \ket{\mathcal{E} + e}_{L} \otimes \ket{\mathcal{E} + e}_{R},
\end{equation}
we can reconstruct the Minkowskian wavefunctional for \( \mathcal{E} = 0 \), given in equation (4.30) of \cite{blomm}, using the Wilson line for an electric field \( \mathcal{E} = \mathcal{E}_{0} \) as
\begin{multline}
  \mathcal{W}_{\mathcal{E}_{0}} \ket{\Psi}
  =
  \mathcal{W}_{\mathcal{E}_{0}} \frac{1}{Z} \sum_{\mathcal{E}} e^{- \frac{\pi R^{2}}{2} \frac{\mathcal{E}^{2}}{2}} \ket{\mathcal{E}}_{L} \otimes \ket{\mathcal{E}}_{R}
  =
  \frac{1}{Z} \sum_{\mathcal{E}} e^{- \frac{\pi R^{2}}{2} \frac{\mathcal{E}^{2}}{2}} \ket{\mathcal{E} + \mathcal{E}_{0}}_{L} \otimes \ket{\mathcal{E} + \mathcal{E}_{0}}_{R} \\
  =
  \frac{1}{Z} \sum_{\mathcal{E}} e^{- \frac{\pi R^{2}}{4} (\mathcal{E} - \mathcal{E}_{0})^{2}} \ket{\mathcal{E}}_{L} \otimes \ket{\mathcal{E}}_{R}.
\end{multline}
On the limit \( R \rightarrow \infty \), we get a delta function \( e^{- \frac{\pi R^{2}}{4} (\mathcal{E} - \mathcal{E}_{0})^{2}} \rightarrow \delta(\mathcal{E} - \mathcal{E}_{0}) \) around \( \mathcal{E}_{0} \) on the exponential and thus obtain
\begin{equation}
  \ket{\mathcal{E}_{0}}
  =
  \ket{\mathcal{E}_{0}}_{L} \otimes \ket{\mathcal{E}_{0}}_{R}.
\end{equation}
Remembering that such constant electric fields are corresponding to $\theta$ originally, we get
\begin{equation}
  \ket{\theta}_{M}
  =
  \ket{\theta}_{L} \otimes \ket{\theta}_{R}.
\end{equation}
The separation of spacetime into right and left wedges correspond to a separation of the state into two sectors, both corresponding to the same parameter \( \theta \).
That result can also be obtained directly from equation (4.31) on \cite{blomm} by applying the Wilson line operator \( \mathcal{W}_{\mathcal{E}} \)
\begin{equation}
  \ket{\mathcal{E}}_{M}
  =
  \mathcal{W}_{\mathcal{E}} \ket{0}_{M}
  =
  \mathcal{W}_{\mathcal{E}} \ket{0}_{L} \otimes \ket{0}_{R}
  \equiv
  \ket{\mathcal{E}}_{L} \otimes \ket{\mathcal{E}}_{R}.
\end{equation}
These results were calculated using a vacuum prepared as an Eigenstate of the field theory in Minkowski space.
Hence, in this particular example (of \( (1 + 1) \)-dimensional pure photodynamics) the $\theta$'s seem to just factorize into right and left sectors, without having to be constrained to the \( \theta = 0 \) case.

As a consequence, take a localized observable \( \mathcal{O} \) with support on the right Rindler wedge.
Then \( \mathcal{O} \) is non-zero only on \( x \in \mathcal{M}_{R} \), and therefore completely accessible to both an inertial and a right-accelerated observer that has \( \mathcal{M}_{R} \) as its causal region.
Then
\begin{equation}
  \label{aveo}
\left< \mathcal{O} \right>_{\Psi}
  =
  \int_{\bfield(\Psi)} [\mathcal{D} A]\,
  \mathcal{O}\, e^{- S_{E}}
  \rightarrow \Tr [\mathcal{O}\, e^{- \beta H_{R}}],
\end{equation}
because the functional integral reduces to a sum over field configurations supported on \( \mathcal{M}_{R} \).
With \( H_{R} \) the Rindler Hamiltonian associated with the $\theta$-dependent action above.

The discussion above is specific to the 1+1 theory because of the relation \cite{adam} between the topological current $\epsilon_{\mu \nu} A^\mu$ and the charge at the horizon $\int dx A_1$.  The charge at the horizon introduces a local Gauss-law constraint  which can be defined and changed locally. As the Reissner-Nordstrom black holes (where the charge is also "local" at the singularity) show, this constraint is valid across horizons.   
In 3+1 dimensions there is no Gauss constraint on $\theta$ and $K_\mu$ is controlled by \equref{windingobs}, which has no equivalent local gauge constraint.

We can make the 1+1 theory a bit more "3D like" if,
in analogy to the previous section \ref{ring} we add the following source terms to the Lagrangian in \equref{lblom}
\begin{equation}
\mathcal{L} \rightarrow \mathcal{L} +  J_\theta \times \left( \theta +J_L(\theta) \right) 
\label{blomsource}
\end{equation}
corresponding to a Rindler-type detector that measures $\theta$ and a device positioned asymptotically at the lest wedge that tunes $\theta$ according to some function (this of course breaks the Gauge constraint).  In this case, we reproduce the same situation as in \equref{gens}.
Equation \ref{aveo} still holds, but the entropy density calculated via \equref{rhodef} from
\[\ s(\beta') \equiv \Tr_R \left[ \left[ \hat{\rho} \ln \hat{\rho} \right]_{\beta'}- \left[ \hat{\rho} \ln \hat{\rho} \right]_{\beta' \rightarrow \infty}  \right]\propto \left( \beta' \right)^{-2} \eqcomma \hat{\rho} = \hat{\rho}_L \times \hat{\rho}_R   \]
will not scale with the same $\beta'^{-1}=a/(2\pi)$ as expected from the Unruh effect but will instead obey an equation such as \ref{gens}.  As a consequence, the ratio of $\beta'$ and $\beta$ is not necessarily set by the microscopic scale,which in this case is the acceleration

Of course, axiomatic quantum field theory \cite{axiom1,axiom2} is not violated here because the source terms $J_\theta,J_L$ introduce a preferred coordinate system which breaks local Lorentz invariance.   However, the effect of topological dynamics in 3D can be argued to be equivalent to a spontaneous generation of $J_\theta,J_L$.  While the scale of $J_\theta$ is local (instanton peak) and of $J_L$ is global (instanton tail), as shown in \equref{gens} scale separation in such a case is not automatically perturbative.  This illustrates that, as argued in section \ref{scales} topological terms can provide a mixing between ``local'' detector-scale physics and infrared scales.

\end{document}